\PassOptionsToPackage{table,dvipsnames}{xcolor}
\documentclass[sigconf, nonacm]{acmart}

\newcommand\vldbpagestyle{plain} 

\providecommand{\eat}[1]{}
\usepackage{subcaption}
\usepackage{cleveref}
\usepackage{multicol}
\usepackage{multirow}
\usepackage{makecell}
\usepackage{tabularx}
\usepackage[table,dvipsnames]{xcolor}

\usepackage[normalem]{ulem} 
\usepackage[linesnumbered,algoruled, lined, noend]{algorithm2e}
\usepackage{enumitem}
\usepackage{bbm} 
\usepackage{calc}
\usepackage{balance}
\usepackage{mdframed}

\SetCommentSty{mycommfont}

\newcommand{\introparagraph}[1]{\medskip \noindent {\bf  #1.}}  

\newmdenv[
  leftmargin=1cm,
  rightmargin=1cm,
  innerleftmargin=0.15cm,
  innerrightmargin=0.15cm,
  innertopmargin=0cm,
  innerbottommargin=0cm
]{customframe}

\newcommand{\insight}[1]{
\noindent\begin{customframe}
    \paragraph{Insights} #1
        \medskip
    \end{customframe}
}

\theoremstyle{definition}

\newtheorem*{assumption*}{\assumptionnumber}
\providecommand{\assumptionnumber}{}
\makeatletter
\newenvironment{assumption}[2]
 {%
  \renewcommand{\assumptionnumber}{Assumption #1-#2}%
  \begin{assumption*}%
  \protected@edef\@currentlabel{#1-#2}%
 }
 {%
  \end{assumption*}
 }

\newtheorem*{property*}{\propertynumber}
\providecommand{\propertynumber}{}
\makeatletter
\newenvironment{property}[2][\unskip]
 {%
  \renewcommand{\propertynumber}{Property #1: #2}%
  \begin{property*}%
  \protected@edef\@currentlabel{#1-#2}%
 }
 {%
  \end{property*}
 }

\definecolor{light-gray}{gray}{0.95}  
\newcommand\regexfmt[1]{\regexfmthelp#1 \relax\relax}
\def\regexfmthelp#1 #2\relax{\allowbreak\grayspace\tokenscolor{#1}\ifx\relax#2\else
 \regexfmthelp#2\relax\fi}
\newcommand\tokenscolor[1]{\colorbox{light-gray}{\textcolor{black}{%
  \ttfamily\mystrut\smash{\detokenize{#1}}}}}
\def\mystrut{\rule[\dimexpr-\dp\strutbox+\fboxsep]{0pt}{%
 \dimexpr\normalbaselineskip-2.25\fboxsep}}
\def\grayspace{\hspace{0pt minus \fboxsep}}

\newcommand{\TODO}[1]{\textcolor{blue}{#1}}
\newcommand{\Ling}[1]{\textcolor{teal}{#1}}
\newcommand{\jmp}[1]{\textcolor{red}{#1}}
\newcommand{\sd}[1]{\textcolor{brown}{SD: #1}}

\newcommand{\free}{\texttt{FREE}} 
\newcommand{\best}{\texttt{BEST}} 
\newcommand{\lpms}{\texttt{LPMS}} 
\newcommand{\vgg}{\texttt{VGGraph}} 
\newcommand{\trigram}{\texttt{Trigrams}} 
\newcommand{\wlweb}{\textbf{Webpages}}
\newcommand{\wldblp}{\textbf{DBLP}}
\newcommand{\wlprotein}{\textbf{Prosite}}
\newcommand{\wltraffic}{\textbf{US-Acc}}
\newcommand{\wlsql}{\textbf{SQL-Srvr}}
\newcommand{\wlsysy}{\textbf{ADX}}
\newcommand{\wlenron}{\textbf{Enron}}
\newcommand{\wlsynnew}{\textbf{Synthetic}}
\newcommand{\wlsyn}{\textbf{Robust}}

\newcommand{\1}[1]{\mathbbm{1}\left[#1\right]}

\newcolumntype{L}{>{\leavevmode}l}
\newcolumntype{C}{>{\leavevmode}c}

\newcolumntype{R}{>{\raggedleft\arraybackslash}X}

\newcommand{\repo}{\url{https://github.com/mush-zhang/RegexIndexComparison}}

\newcommand{\freePlan}{\Cref{app:free_plan}}
\newcommand{\robustTest}{\Cref{app:robustness}}

\begin{document}
\pagenumbering{arabic}
\pagestyle{\vldbpagestyle}

\title{An Evaluation of N-Gram Selection Strategies for Regular Expression Indexing in Contemporary Text Analysis Tasks. Extended Version
}

\author{Ling Zhang}
\affiliation{%
    \institution{University of Wisconsin-Madison}
    \city{Madison}
    \state{WI}
    \country{United States}
}
\email{ling-zhang@cs.wisc.edu}

\author{Shaleen Deep}
\affiliation{%
    \institution{Microsoft Gray Systems Lab}
    \country{United States}
}
\email{shaleen.deep@microsoft.com}

\author{Jignesh M. Patel}
\affiliation{%
    \institution{Carnegie Mellon University}
    \city{Pittsburgh}
    \state{PA}
    \country{United States}
}
\email{jignesh@cmu.edu}

\author{Karthikeyan Sankaralingam}
\affiliation{%
    \institution{University of Wisconsin-Madison}
    \city{Madison}
    \state{WI}
    \country{United States}
}
\email{karu@cs.wisc.edu}

\renewcommand{\shortauthors}{Zhang et al.}

\begin{abstract}
Efficient evaluation of regular expressions (regex, for short) is crucial for text analysis, and n-gram indexes are fundamental to achieving fast regex evaluation performance. However, these indexes face scalability challenges because of the exponential number of possible n-grams that must be indexed. Many existing selection strategies, developed decades ago, have not been rigorously evaluated on contemporary large-scale workloads and lack comprehensive performance comparisons. Therefore, a unified and comprehensive evaluation framework is necessary to compare these methods under the same experimental settings. 
This paper presents the first systematic evaluation of three representative n-gram selection strategies across five workloads, including real-time production logs and genomic sequence analysis. We examine their trade-offs in terms of index construction time, storage overhead, false positive rates, and end-to-end query performance. Through
empirical results, this study provides a modern perspective on existing n-gram based regular expression evaluation methods, extensive observations, valuable discoveries, and an adaptable testing framework to guide future research in this domain. We make our implementations of these methods and our test framework available as open-source at \repo. \eat{\jmp{the link is not currently active/open} \Ling{just made public}}
\eat{Our experiments reveal that lightweight heuristic-based methods outperform computationally intensive alternatives, reducing index build time and space overhead by more than 90\% compared to the exhaustive selection, with only around a 1\% increase in workload execution time.} 
\end{abstract}

\maketitle
\everypar{\looseness=-1}


\section{Introduction}
Regular expressions are a foundational tool for text pattern matching, powering critical applications such as real-time log analysis~\cite{Rouillard_2004}, genomic sequence alignment~\cite{Arslan_2005}, and web information retrieval~\cite{De_Bra_1994}. However, as datasets grow in their size, the computational cost of brute-force matching becomes prohibitive. To address this issue, n-gram indexing has been widely adopted to accelerate regex processing by pre-filtering candidate text regions using selected n-grams~\cite{Ogawa_1998, cox2012regular}. Despite its widespread adoption, the scalability of this approach hinges on a critical problem: how to select the optimal set of n-grams to index, balancing trade-offs between index size, construction time, and query accuracy.

\eat{Existing solutions, such as frequency-based selection (prioritizing common n-grams)~\cite{FREE}, coverage-optimized methods (minimizing redundant matches)~\cite{BEST}, and heuristic-driven strategies like \lpms~\cite{FAST}, were proposed decades ago. For instance, the \free~\cite{FREE} and \best~\cite{BEST} methods, introduced in the early 2000s, focus on minimizing computational space overhead and index size for memory-constrained systems. Despite being an active area of research, surprisingly, the performance of these methods on modern hardware is lacking for three reasons. {The absence of a comprehensive, unbiased, and systematic understanding of these methods is due to three main reasons.} First, existing methods have not been evaluated within the same experimental settings{ (such as the same hardware or using the same B+-Tree implementation)}, leading to incomplete comparisons. Second, prior work has only compared specific metrics and on a limited number of datasets. Third, the absence of a standardized and inclusive testing framework hinders the investigation and testing of methods in practical implementations. Consequently, the lack of a modern comparison of these past methods has led to the adoption of these legacy strategies in recent studies~\cite{Qiu_2022, Chockchowwat_2022} without re-evaluating their assumptions. These older methods may not be appropriate when dealing with larger datasets (now common in applications domains such as genenomics, IoT/sensor, and log analytics) as it stresses their n-gram selection methods. For example, as we will see later, \best’s quadratic runtime complexity becomes prohibitive on even a small 10GB text corpus, forcing practitioners to downsample data or abandon indexing altogether. In fact, n-gram indexing used by Postgres~\cite{postgres_trg} and GitLab code search~\cite{gitlab} abandon the n-gram selection strategies and build the index with all n-grams for only trigrams (i.e. $n=3$) or for several fixed values $n$ (for instance, indexing all bigrams and trigrams). In summary, there are three key limitations that motivate our work.}

Legacy methods like frequency-based selection~\cite{FREE}, coverage-optimized strategies~\cite{BEST}, and heuristics like \lpms~\cite{FAST} were introduced decades ago to reduce index size and memory usage~\cite{FREE,BEST}. Despite being an active area of research, surprisingly, the performance of these methods on modern hardware is lacking for three reasons. {The absence of a comprehensive, unbiased, and systematic understanding of these methods is due to three main reasons.}: (1) inconsistent experimental setups hinder fair comparisons, (2) evaluations are narrow in scope and datasets, and (3) a lack of standardized testing framework hindering practical assessment. As a result, recent studies~\cite{Qiu_2022, Chockchowwat_2022} continue to adopt these outdated approaches without re-evaluating their assumptions. These methods struggle with today’s large datasets (e.g., genomics, IoT, logs), where selection strategies break down. For instance, \best’s quadratic runtime becomes impractical on even 10GB of text, often forcing downsampling or skipping indexing. Modern systems like Postgres~\cite{postgres_trg} and GitLab~\cite{gitlab} bypass selection entirely, indexing all trigrams or fixed n-grams. These limitations motivate our work.

\begin{enumerate}[label=\textbf{M.\arabic*}, ref=\textbf{(M.\arabic*)}, left=0pt, wide]
\item \textbf{Lack of Comprehensive Comparison.} Prior evaluations focus on specific metrics (e.g., false-positive rate) or evaluate their method using only a single workload, failing to capture the diverse demands of real-world workloads. \eat{For example, genomic databases prioritize minimizing false positives due to a large number of repeated queries on the same dataset, while log analysis prioritizes rapid index rebuild overhead for large scale streaming data analysis.} No study has systematically compared and contrasted the three key methods \free, \best, and \lpms\ across wide-ranging scenarios. \label{m1}

\item \textbf{Outdated Resource Assumptions.} Existing methods often optimize assuming limited memory, investing significant efforts to reduce memory overhead and index sizes. \eat{Therefore, a lot of effort is invested in reducing the memory overhead during computation and reducing the index sizes. }Additionally, some approaches focus on scenarios where the index and/or dataset cannot fit in memory and use the IO cost as an optimization parameter~\cite{Kim_2010}. However, in many modern hardware settings, abundant memory is often available, and it is also important to consider this case.\eat{cases where the entire dataset and index fit in memory. This shift allows us to examine whether existing methods can leverage the increased memory capacity to enhance performance.} \label{m2} 

\item \textbf{Fine-grained empirical analysis.} Beyond evaluating the existing methods on a comprehensive set of common metrics, we also conduct detailed resource usage measurements for each method, providing deeper performance related insights. \eat{This systematic approach ensures an unbiased and systematic comparison, providing deeper insights into their performance and efficiency.} \label{m3}

\end{enumerate}

\introparagraph{\textit{Our Contributions}}  
We systematically study three state-of-the-art methods, \free,\ \best,\ and \lpms,\ across various workloads. \eat{By implementing these strategies with parallelization and evaluating them on diverse datasets, we provide a fair comparison of their performance, quantify their trade-offs, and test their scalability on larger datasets.} Specifically, we make the following contributions.

\introparagraph{1. Systematic Method Analysis} We formalize and categorize three state-of-the-art n-gram selection strategies, \free\ (frequency-based), \best\ (coverage-optimized), and \lpms\ (linear programming approximation), within a unified framework. This enables direct theoretical comparison of their computational complexities. Additionally, we provide a detailed account of their implementation and design decisions, emphasizing both similarities and subtle differences with other methods, addressing~\ref{m1}

\introparagraph{2. Methods and Benchmark Framework Implementation} In the absence of original code, we implemented \free, \best, and \lpms\ as described in the respective papers, using modern C++. While \best’s original design includes parallelism for multi-core CPUs, \free\ and \lpms\ lack scalable implementations. We modernized all three methods by: (1) redesigning \free\ with parallelism to leverage modern multi-core CPUs, and (2) implementing \lpms\ with Gurobi, which inherently supports multi-threading in its LP solver. \eat{Additionally, we introduced an early stopping mechanism to all three methods to configure the maximum number of n-grams selected. We also developed an end-to-end benchmark framework that manages data loading, method selection, configuration settings, regex matching, and result reporting.} Our benchmarking framework is modular, easily extensible, and publicly available at the link noted in the abstract. All experiments were conducted on a multi-core machine with large memory, addressing~\ref{m2}.

\introparagraph{3. Broad Workload Benchmark} To address the lack of comprehensive empirical comparisons on workloads with different characteristics, we evaluate the methods across {eight} workloads. This includes using three legacy benchmarks from prior work and two real-world workloads in contemporary regex matching tasks. 

\introparagraph{4. Empirical Trade-off Analysis and Guidelines} Our experiments quantify performance across various workloads by measuring index construction time, memory footprint, and index precision, among other metrics (addressing~\ref{m3}). We correlate these results with workload characteristics, such as regex complexity and dataset size, to derive actionable guidelines. For instance, \free\ achieves a 92\% reduction in index build time compared to \best\ with only a 1.2\% increase in query latency, making it more suitable for streaming log analysis. These findings challenge the necessity of computationally intensive methods for large-scale workloads.



\eat{
\begin{enumerate}
\item Systematic discussion of the key methods: FREE, BEST, and FAST. We explained the logics and assumptions  of these methods under the same set of theoretical definitions. We also 
\item A holistic implementation of these methods to allow an apple-to-apples comparison. Including multi-threaded implementation, which in some cases had not been implemented before. In a modern hardware setting, studying performance in a multi-threaded setting is critical for practical consideration.
\item Creating a broad collection of datasets that has been used in the past and adding new real-life datasets to create a broader dataset for evaluation, which has been missing in the past studies making it hard to understand the behavior of these methods in diverse data settings.
\item An empirical evaluation of these methods that clearly shows the performance benefits of each approach. 
\end{enumerate}
}

\section{Related Work}
Efficient indexing of regular expressions is crucial to enhance query performance in large-scale database systems, log analysis, and information extraction applications. Traditionally, regular expressions are evaluated using finite automata, which is computationally expensive due to backtracking and state transitions. Since regex queries operate on string datasets, n-gram-based indexing techniques have been widely adopted to accelerate regex evaluation by pre-filtering candidate matches before full regex matching. N-gram selection techniques are critical for regex indexing, as they impact both query filtering accuracy and index storage overhead.

To maximize regex evaluation performance before the introduction of n-gram indexing techniques or when such indexing is not feasible, early research focused on string literal filtering to optimize regex performance. A suffix-based string matching heuristic was presented in~\cite{Boyer_1977}, laying the foundation for later suffix-tree-based regex indexing methods. Suffix-based indexing was then proposed for efficient pattern matching over large datasets~\cite{suffix_tree_1997}. Literal filtering, or prefix heuristics for approximate string matching, was introduced in~\cite{Wu_1992, Clarke_1996}. These studies demonstrated that prefix-based early termination can significantly reduce regex search time. Later works proposed filtering with all discriminative string literals in addition to prefixes and suffixes in the regexes~\cite{hyperscan, Zhang_2023}.

N-gram indexing was introduced to further improve regex query performance by precomputing substrings of length n as index keys. Besides being used for pattern matching, n-gram indexing is also common in approximate string searching. Early works compared different n values and decided to index all trigrams. {We refer to this method as \trigram, and will use it for comparison in~\Cref{sec:expr}.} Other works index n-grams with different n values for various workloads. The database community has conducted extensive research on effective indexing strategies for regular expression queries. While existing works primarily focus on improving regex query performance through optimized index structures, relatively fewer studies have explored n-gram key selection strategies for regex indexing. N-gram indexing has also been used for the closely related problem of indexing regular expressions (instead of data) to find out which regexes match a given input string~\cite{chan2003re}. Similar ideas have also been used for indexing data to speed up regular path query evaluation~\cite{li2001indexing}.

The first known work discussing n-gram selection  for regex indexing is \free~\cite{FREE}. The high-level idea of selecting a covering n-gram set was proposed in a study of Asian language indexing~\cite{Ogawa_1998}. The concept of a covering set was further developed by Kim et al.~\cite{Kim_2010} for a more efficient n-gram selection method considering I/O cost. The analogy between the set-covering problem and the n-gram selection problem was formalized in \best~\cite{BEST}, where the authors presented a near-optimal algorithm to select variable-length n-grams considering index size constraints. Later work, \lpms~\cite{FAST}, combined the solution formulations of \free\ and \best\ and reformulated the n-gram selection problem into linear programs. In this work, we compare the n-gram selection and indexing methods of \free, \best, and \lpms\ for in-memory workloads and indices. 

{Several works have explored ML-based approaches for regex evaluation in security contexts. On resource-constrained edge devices, compressed on-device ML models are commonly used for high-throughput malicious packet detection. However, these methods rely on extensive preprocessing and assume access to high-quality training data—assumptions that don’t hold in log analytics, where analysis is routinely ad hoc. For a broader overview, see the survey by Xu et al.~\cite{xu2016survey}.}


\section{Problem Definition} \label{sec:prelim:problem}

In this section, we discuss the n-gram selection problem and provide formal definitions for some common terms.

\begin{definition}[N-Gram]
For a given finite alphabet $\Sigma$, an n-gram $g$ is a sequence of $n$ characters $g = g_1g_2\cdots g_n$ where $g_i \in \Sigma$.
\end{definition}

We use $|g|$ to denote the length of the n-gram $g$. A literal is a string from the set $\Sigma*$, where $*$ is the standard closure operator. A common operation on a regex query $q$ is to identify the maximal literal components from the regex. For a literal $l$, we will use $G(l)$ to denote the set of all possible substrings of $l$. The set of all possible n-grams of a regex $q$ is $G(q) = \bigcup_{\text{maximal }l \in q} G(l)$.
{We use $G$ to define the set of all candidate n-grams considered in an n-gram selection method. Note that $G$ may be different for different methods, and we will further discuss it in~\Cref{sec:methods}. $G_i$ to is used to denote subset of candidate n-grams that has length exactly $i$.}

\eat{Different from the common definition of n-gram in natural language processing sense, where the smallest token of analysis is a \textit{word}, the smallest token in regex queries is usually a \textit{character}. Regex queries often times look for patterns inside a \textit{word}, for example, one regex \regexfmt{<a href=("|').*ZZZ\.pdf("|')>} from a workload of regex queries over webpages matches pattern that is itself a word in term of a URL, and therefore having token of analysis as large as \textit{word} will not help in this regex.}
\eat{\sd{Make this a proper example using begin\{example\}. Also, use the highlighting for literals like we did in the BLARE paper.}\Ling{ Ling: added}}
\begin{example}
Regex queries often times look for patterns as a sequence of characters. The regex \regexfmt{<a href=("|').*ZZZ\.pdf}\regexfmt{("|')>} from a workload of regex queries over webpages matches a URL pattern. The set of literal components in the example regex are:
\{\regexfmt{<a href=}, \regexfmt{ZZZ.pdf}, \regexfmt{>}\}. 
\end{example}

A workload $W = (Q,D)$ consists of the set of regex {queries} $Q=\{q_1, q_2, \cdots\}$ and dataset $D=\{d_1, d_2, \cdots\}, d_i \in \Sigma*$. The size of the dataset is defined as $|D| = \sum_{d_i \in D} |d_i|$. Each $d_i$ will be referred to as a data record. Next, we define the support of an n-gram.
\begin{definition}[Support]
    The support $s$ of a literal $g$ in dataset $D$ is the number of elements in $D$ that contains $g$. Similarly, its support in query set $Q$ is the number of individual queries which contains $g$ as a literal.
    $$s_D(g) = \sum_{d\in D} \1{g \in G(d)} \quad s_Q(g) = \sum_{q\in D} \1{g \in G(q)} $$
\end{definition}

Using the notion of support, we define the selectivity of an n-gram $g$ as follows.

\begin{definition}[Selectivity]
    The selectivity $c$ of an n-gram $g$ in string set $D$ is the fraction of individual strings in $D$ that contains $g$, i.e., $c_g = {s_D(g)}/{|D|}$.
\end{definition}

For n-gram selection methods, selectivity is an important parameter when deciding if an n-gram is selected for indexing or not. For the previous example regex that looks for a URL of a PDF file which has a filename end with sub-string \regexfmt{ZZZ}, by looking for strings with n-grams \regexfmt{ZZZ.pdf}, intuitively, we can reduce the number of data strings for the exact regex matching. Notice that there are two other literals in the regex, \regexfmt{<a href=} and \regexfmt{>}. Since these two literals almost always occur in every webpage HTML, indexing these two literals might not reduce the number of data records for exact regex matching, that is, high selectivity. Further, it may result in extra computation overhead. Conversely, n-gram that has lower selectivity like \regexfmt{ZZZ.pdf} can help reduce the overall query time.
\eat{\sd{Again, highlight the literals using background coloring like the BLARE paper. It looks better.}\Ling{Added}}

However, lower selectivity is not always better, when considering the entire query set, Q. For instance, the n-gram \regexfmt{ZZZ.pdf} may have low selectivity and effectively filter out irrelevant data points. Since \regexfmt{ZZZ} is an uncommon character sequence in the English language, it might not appear in other queries. On the other hand, indexing \regexfmt{.pdf}, which has relatively higher selectivity than \regexfmt{ZZZ.pdf}, might not reduce as many data points, but it can benefit other queries that look for URLs of PDF files with different names. Thus, there is a trade-off between selecting higher and lower selectivity n-grams.

\section{Methods Overview} \label{sec:methods}
In this section, we present the overview of the three state-of-the-art n-gram selection methods: \free, \best, and \lpms. We will describe their selection strategy and provide a complexity analysis. In~\Cref{tab:method_summary}, we summarize and compare the three methods in terms of their source of n-gram, selection criteria in each step, index structure accompanied, and other common configurations.

\begin{table*}[t]
\centering
\captionof{table}{Selection methods summary.}

\eat{  
}

\scalebox{0.95}{
\small
\begin{tabularx}{1.05\linewidth}{ L | LLLLLL}
\toprule
 Method & \makecell{Source} & \makecell{Selection \\Criteria} & \makecell{Selectivity \\Threshold ($c$)} & \makecell{N-Gram \\Constraint} 
 & Time Complexity & Space Complexity\\ \midrule
\free~\cite{FREE} & \eat{dataset}{$D$} & \makecell[l]{Prefix-free\\Selectivity} & 0.1 & $[2, 10]$ & {$O(k \cdot |D|)$} & {$O(|D|)$}\\\midrule
\best~\cite{BEST}  & \eat{queries \& dataset}{$Q,D$} & Utility & $[0.05, 0.1] $ & - & {$O(|D|+|Q|)$} & {$O(|I| \cdot |G(Q)| \cdot |D| \cdot |Q|)$}\\\midrule
\lpms~\cite{FAST} & \eat{queries \& dataset}{$Q,D$} & \makecell[l]{Prefix-free\\Utility optimized} & - & - & \makecell[l]{{$O(\sum_{i \in [10]}\{|G_i(Q)|^{2.5} $}\\ {$\quad + |G_i(Q)| \cdot (|D| + |Q|)\})$}} & {$O(\max_{i \in [10]} |G_i(Q)|\cdot(|D|+|Q|)$} \\\midrule
\vgg~\cite{Qiu_2025} & {$Q,D$} & Combines \free\ and \best & - & $[2,\inf)$ & {$O(n \cdot |D| + |Q|\cdot|G(Q)|\cdot n)$} & {$O(|D|+|G(Q)|)$\eat{$O(|I| \cdot |G| \cdot |D| \cdot |Q|)$}}\\\midrule

\end{tabularx}}
\label{tab:method_summary}
\end{table*}

\subsection{FREE}\label{subsec:methods:free}

\subsubsection{Selection Strategy}
\free\ uses the dataset in the workload as the source of n-gram selection, and selects a prefix-free set of n-grams based on selectivity. Note that \free\ does not use the query workload for n-gram selection. The candidate n-gram set {$G$} is $G(D) = \bigcup_{d \in D} G(d)$.
\eat{For the candidates $G$, }\free\ decides whether {an n-gram} will be selected by the \textit{Usefulness} criteria as defined below.

\begin{definition}[\free\ Usefulness]
By setting a fixed \textit{selectivity threshold} $\mathbf{c}$, n-gram $g$ is deemed \textit{useful} if its selectivity is less than the \textit{selectivity threshold}, $c_g < \mathbf{c}$. 
\end{definition}

\free\ makes two important assumptions for n-grams selection.

\begin{assumption}{\free}{1}\label{assumption:free:1}
\free\ assumes that n-grams with high selectivity are less \textit{useful}.
\end{assumption}

\begin{assumption}{\free}{2}
With a careful selected \textit{selectivity threshold} $\mathbf{c}$, \free\ assumes that a query without any \textit{useful} n-grams is rare. Overall workload performance will still improve as most of the regexes can benefit from indexing only \textit{useful} n-grams.
\end{assumption}

\free\ selects only n-grams that are \textit{useful}. Since \textit{usefulness} selection criteria is essentially a selectivity upper bound, if we know that {an n-gram} $g$ is \textit{useful}, then any longer n-gram having $g$ as a substring is also useful. For example, if n-gram \regexfmt{pdf} is \textit{useful} with a selectivity $c_g$, then n-grams such as $g' = $ \regexfmt{.pdf} and $g'' = $ \regexfmt{pdfg}\eat{ \sd{background color literals}} will have selectivity \eat{\$0 \leq }$c(g') < \mathbf{c}$ and \eat{\$0 \leq }$c(g'') < \mathbf{c}$, therefore also \textit{useful}. 

\begin{property}[\free]{Usefulness}\label{thm:free:extend}
For a {\textit{useful}} n-gram $g = g_1g_2\cdots g_{n_1}$ of size $n_1$ having selectivity $c_g$, any other longer n-grams of size $n_2$ $$g' = p_1\cdots p_{m}g_1g_2\cdots g_{n_1}s_1\cdots s_{n_2-n_1-m}$$ with a prefix size $m$ and suffix size $n_2-n_1-m$, where each character $p_i, s_i\in \Sigma$, we have $0\leq c_{g'} \leq c_g$, is also \textit{useful}.
\end{property}

Since the set of all \textit{useful} n-grams can still be too large, to further reduce the index size, \free\ selects only the minimal n-gram among all n-grams with the same prefix. For example, n-grams \regexfmt{pdf} and \regexfmt{pdfg} have the same prefixes \regexfmt{p}, \regexfmt{pd}, and \regexfmt{pdf}. If the selectivity of the first two n-grams are not \textit{useful} while \regexfmt{pdf} is \textit{useful}, we index only n-gram \regexfmt{pdf}, although \regexfmt{pdfg} is also {useful}. Thus, the second n-gram selection criterion besides selectivity is \textit{minimality}.

\begin{definition}[\free]{Minimal}
\looseness=-1 For an n-gram $g$ in alphabet $\Sigma$ of size $n$ that is \textit{useful}, it is \textit{prefix-minimal} among the set of n-grams that has $g$ as a prefix if no prefix substring of $g$ with size smaller than $n$ is \textit{useful}. An n-gram $g$ is \textit{suffix minimal} if no suffix substring of $g$ with size smaller than $n$ is \textit{useful}.
N-gram $g$ is \textit{prefix-suffix minimal}, or \textit{pre-suf minimal} if no substring of $g$ with size smaller than $n$ is \textit{useful}.
\end{definition}

To effectively reduce the size of n-grams selected, \free\ selects the prefix-minimal useful set of n-grams, deriving from the Apriori method of finding the maximal frequent sets in data mining literature~\cite{agrawal_1996}. Essentially, it generates all candidate n-grams in increasing size order iteratively. For iteration $i$, it generates the candidate set of n-grams of length $i$ by extending all useless n-grams of length $i-1$ by one character, inserting the useful n-grams in the candidate set into the index, and using the useless ones in the candidate set for iteration $i+1$. This way, it is not necessary to generate all possible n-grams in each iteration. This method also ensures that the set of n-grams in the index is a prefix-minimal set, as the breadth-first search ensures that the shortest prefix of a useful n-gram is visited first. No n-grams in candidate sets of future iterations have the selected n-grams as prefixes, as the useful n-grams are never extended to generate a candidate n-gram.

\free\ confines the search space by constraining the length $n$ of the n-grams selected. Briefly, there are two parameters to tune: 1) selectivity threshold $c$ that distinguish {useful} n-grams from {useless} n-grams, and 2) n-gram size $n$ to control the length of the index keys. In the original paper, the authors uses $c=0.1$ and $2\leq n \leq 10$ in their experiments. {It also defined a simple query plan considering the index keys. We include a description of the query plan generation in \freePlan\ of the full paper. We implemented the original version of \free\ with the query plan and an accompanied matcher in the codebase, but for a fair comparison of the n-gram selection methods, we do not use it for the experiments.}
The index data structure used is inverted index with n-grams as keys and posting lists as values.

\subsubsection{Complexity Analysis} \label{subsec:free:complexity}
Let the candidate set at step $i$ be $M_i$. Initially, \free\ scans the dataset to identify all unique unigrams and compute their selectivities, using a temporary hashmap of size $|M_1| = |\Sigma|$. This base step has time complexity $O(|D| + |\Sigma|)$. After filtering for useful unigrams, the remaining ones are extended to form bigrams $M_2$, with space and compute overheads of $O(|M_2|)$ and $O(|D| + |M_2|)$, respectively.

At each step $i$, the runtime space is $O(|M_i|)$ and compute time is $O(|D| + |M_i|)$. Since previous hashmaps are discarded, selecting n-grams up to size $k$ results in total space $O(|M_k|)$ and compute time $O(k \cdot (|D| + |M_k|))$. Given that $\sum_{i=1}^k |M_i| \leq |D|$, the overall space is $O(|D|)$ and compute time is $O(k \cdot |D|)$. {In practice, $k$ is usually small and can be treated as a constant.}

\subsection{BEST}\label{subsec:methods:best}
\looseness=-1 Although \free\ is effective in selecting n-grams with high filtering power for indexing, it has the limitation of not considering the query set. As a result, it cannot guarantee that the index will be as helpful when the character frequency distribution for the query literals differs from that of the dataset. \best\ remedies this problem.

\subsubsection{Selection Strategy} \best\ utilizes both the dataset and the query set as sources for n-gram selection. In addition to considering the \textit{selectivity} of n-grams within the dataset, it also takes into account the frequency of n-gram occurrences in the query set. This approach helps avoid selecting n-grams that do not benefit any regex query, as illustrated in the example in~\Cref{sec:prelim:problem}.

\begin{assumption}{\best}{1}
    It is \textit{beneficial} to index an n-gram that \textbf{does not appear} in a data record, so that it can filter out the data earlier; it is \textit{beneficial} to index an n-gram that \textbf{appears} in a query, so that it can be used to filter out data records for the query.
\end{assumption}

\begin{assumption}{\best}{2}\label{assumption:best:2}
    {The benefit of filtering out a data record $d_i$ is similar to the benefit of filtering out a data record $d_j$, where $i\neq j$.}
\end{assumption}
With these assumptions, \best\ abstracts the n-gram selection problem into a graph cover problem. Each data record, query, and n-gram is regarded as an individual node in the universe $U$. If an n-gram $g$ is present in a query $q$, there is an edge between $g$ and $q$; if $g$ is absent from an input string $d$, there is an edge between $g$ and $d$. In the context of our regex workload, each regex is matched once, so all edges have equal weights due to Assumption~\ref{assumption:best:2}. Each subset in $U$ represents a set of connected nodes interlinked by a single n-gram. 
Formally, we have the following construction.
\begin{definition}[\best\ Cover]\label{def:best:cover}
    For a workload $W = (Q,D)$, the cover of an n-gram $g$ is the set of data records $d \in D$ and query $q \in Q$ pairs defined as follows:    
    {$$cover(g)=\{(q,d)\in Q\times D\mid g\in q\land g\notin d\}$$}
    
    For a set of n-grams $P$, the cover of the n-gram set is the union of the cover of each n-gram in the set.
    $$cover(P) = \bigcup_{g \in P} cover(g)$$
    
\end{definition}

The n-gram selection problem is then transformed to a budgeted maximum set cover problem~\cite{Khuller_1999} that aims to find the {set that maximize the cover over all sets $P$ that satisfy the budget constraint}\eat{minimum set $P$ that achieves maximum cover.} 
Formally, \best\ define the value of indexing an n-gram with respective to the entire workload by its \textit{benefit}. 

\begin{definition}[\best\ Benefit]
    For a workload $W = (Q,D)$ and an n-gram set $I$, the benefit of an n-gram $g \notin I$ is the number of additional query-data pairs covered by $g$ that is not already covered by $I$. 
    $$benefit(g,I)= |cover(I\cup \{g\}) - cover(I)|$$
\end{definition}
{To progressively calculate the benefit of each n-gram $g$ of $G(W) = G(D) \bigcup G(Q)$, we need two matrices to store the existence of $g$ in each data record $d$ and each query $q$. Specifically, the two matrices are of size $|Q|\cdot |G(W)|$ and $|D| \cdot |G(D)|$ respectively.}
\begin{definition}[\best\ Cost]
    For a workload $W$ consists of dataset $D$ and query set $Q$, the cost to index an n-gram $g$ is the storage overhead of $g$. 
\end{definition}
The {cost} is dependent of the index structure. In the original \best\ paper, B+-tree is used, and the cost of $g$ is the number of leaf pointers corresponding to $g$, which number of data records in $D$ that contains $g$. For inverted index, the cost of $g$ is the size of its list, which is of size $s_D(g)$.
\begin{definition}[\best\ Utility]
    For a workload $W$ consists of dataset $D$ and query set $Q$  and an existing n-gram set $I$ on $W$, the utility of indexing an additional n-gram $g$ is the ratio of its benefit over cost, i.e.,
    $utility(g) = {benefit(g,I)}/{cost(g)} $
\end{definition}
\best\ selects n-grams based on their {utility}. The brute-force method would be to iteratively select the n-gram with the highest {utility}, $g_{max}$, based on the workload $W$ and the set of selected n-grams $I$, among all possible n-grams with a positive benefit.
\eat{We notice that in order to compute the \textit{benefit} and \textit{utility} of an individual n-gram $g$, we need to know the existence of $g$ in each query $q\in Q$ and data record $d\in D$ according to Definition~\ref{def:best:cover}.}

\begin{assumption}{\best}{3}\label{assumption:best:3}
    {The average selectivity of candidate n-grams is low for the data records and the literals in the queries.}
\end{assumption}
\begin{assumption}{\best}{4}\label{assumption:best:4}
    The average number of characters in all literals of a regex is much smaller than that of data records. The number of regex queries is much smaller than number of data records in the same workload.
\end{assumption}
Therefore, according to the sparsity assumption~\ref{assumption:best:3}{ and the definition of cover}, {instead of considering all n-grams in the query set and dataset, we can trim away the n-grams that only exists in the dataset. Therefore, the candidate n-gram set $G$ of \best\ is $G(Q)$.}
\best\ choose to use adjacency lists: \textit{$Q$-$G$-list} and \textit{$G$-$D$-list} rather than two matrices of sizes $|Q|\cdot|G(W)|$ and $|G(W)|\cdot |D|$ to store the existence of {candidate n-grams}\eat{n-grams in the workload} to reduce space usage. The candidate n-grams, queries, and data records are each assigned an index number. The indices of \textit{$Q$-$G$-list} correspond to query numbers, and each element is a list of n-gram numbers in this query. Similarly, the indices of \textit{$G$-$D$-list} correspond to n-gram numbers, and each element is a list of data record numbers which the n-gram is in. 

\subsubsection{Approximation Techniques} \label{method:best:approximate}
The budgeted maximum set cover problem optimization problem is NP-hard~\cite{Lund_1994,Khuller_1999}. 
Besides, the search space is the product of number of possible query-data pairs, $|Q|\cdot |D| $ and the number of possible n-grams, $|G|$. \eat{ By Assumption~\ref{assumption:best:4}, we assume $G = G(Q)$. }When the query size and the dataset size is large, the number of candidate {query-data pairs}\eat{n-grams} can be prohibitively large to select the n-grams using brute-force method.
\best\ employs several techniques to get an approximate result.

\introparagraph{Pruning} \best\ introduce pruning of n-grams by {selectivity} to reduce the number of the candidate n-grams. With the same assumption as Assumption~\ref{assumption:free:1}, \best\ prune n-grams that has {selectivity} larger than a threshold $c$. 
  
\introparagraph{Parallelism by Clustering} \best\ clusters the regex queries into small groups of similar queries that contains largely overlapping n-grams. Within each small group $Q_i$, the computation will be $O(|Q_i|\cdot |D| \cdot |G(Q_i)|$. Formally, the distance is calculated as
$$Dist(q_1, q_2) = \frac{|(G(q_1) -G(q_2)) \cup (G(q_2) - G(q_1))|}{|G(q_1) \cap G(q_2)| }$$ Since queries with similar set of n-grams are clustered together, we minimize the number of n-grams for each subproblem, that is, the size of $G(Q_i)$.
The clustering allows \best\ to divide the search in each iteration into smaller sub-problems that allows for parallel computation. The intermediate {cost} and {benefit} results from all sub-problems are aggregated at the end of each iteration to select the n-gram with maximum {utility}.

\introparagraph{Workload Reduction} When the size of the query set $Q$ is large, \best\ selects a representative sample $Q' \subseteq Q$ to further reduce the computation overhead. To ensure that the sample is representative, \best\ use the same clustering technique to cluster the literals in $Q$ into clusters. When the clusters stabilize, the median query of each group is selected into $Q'$.

\subsubsection{Complexity Analysis}\label{subsec:best:complexity}

First, the workload reduction step shrinks the query set from $|Q|$ to $|Q'| = \frac{|Q|}{t}$, with the candidate n-gram set $G'$ also reduced proportionally. Assuming average query length $\mu$ and $T$ iterations of k-median, this step takes $O(|Q| \cdot |Q'| \cdot \mu \cdot T)$ time. A suffix tree is used to enumerate all n-grams in $O(|G'|)$ time. 
Next, clustering enables parallelism and, per the original \best\ paper, reduces time and space by $\sim$5 $\times$ even on a single thread. Using adjacency lists (\textit{Q-G-list} and \textit{G-D-list}) instead of matrices cuts space from $O(|G| \cdot (|D| + |Q|))$ to $O(\mu \cdot (|D| + |Q|))$, which is small since $\mu \ll |G|$. However, building these lists still takes $O(|G| \cdot (|D| + |Q|))$ time. By Assumption~\ref{assumption:best:4}, $\mu \ll |G|$, and therefore, the memory overhead during computation is small. Note that we still need $O(|G|\cdot (|D|+|Q|))$ time to build both data structures. The two adjacency lists contributes to the majority of extra space overhead for \best. In each iteration, \best\ will examine all remaining candidate n-grams and their benefit considering the index $I$ built so far. For each pair $(q, d) \in (Q' \times D)$, it checks if the pair covered by each candidate n-gram $g$. In each iteration, \best\ evaluates all remaining n-grams against $(q, d) \in Q' \times D$, costing {$O(|I| \cdot |G(Q)| \cdot |Q| \cdot |D|)$}.

\subsection{LPMS}\label{subsec:methods:lpms}
Despite several optimizations in the \best\ algorithm, its complexity analysis is still too large. \lpms\ remedies this issue by introducing approximations in the algorithm for \best\ via integer programming.

\subsubsection{Selection Strategy} 
Similar to \best, \lpms\ also uses both the query set and the dataset as sources of n-gram selection. \lpms\ also incorporates Assumption~\ref{assumption:free:1} that an n-gram that eliminates more data records is more \textit{useful}, but it incorporates the impact of queries by adjusting it with the \textit{selectivity} of n-grams in queries and the length of the n-gram. Formally,
\begin{definition}[\lpms\ Coverage]
    The \textit{coverage} of an n-gram $g$ is defined as the ratio of the support of $g$ in the dataset $D$ and the support of $g$ in the query set $Q$, normalized by the n-gram length: 
    $cv(g) = {s_D(g)}/{ |g| \cdot s_Q(g) }$
\end{definition}
Similar to Assumption~\ref{assumption:best:4}, we assume $G = G(Q)$. Using the binary variable $x_g = \1{g \in I}$, we form the objective function as $\sum_{g\in G} cv(g)x_g$. Note that $x_g \in \{0, 1\}$ $ \forall g \in G$.
To provide good approximation to the optimal solution, the approximate set of n-grams should satisfy the following:
\eat{\begin{assumption}{\lpms}{1.1}\label{assumption:lpms:1.1}
For each query, there should be at least one n-gram\eat{ in the query indexed}. \sd{Do you mean just index? what is query indexed?}
\end{assumption}}
\begin{assumption}{\lpms}{1}\label{assumption:lpms:1}
The index should filter out at least as many data records {compared to} any candidate n-gram. \eat{\sd{This needs to be reworded. I don't understand this.}\Ling{removed previous assumption and updated this. Basically means that the after filtering with the index (set of n-grams), there should be less or equal number of data records for full regex eval, compared to filtering with any one of the n-grams in the candidate set.}}
\end{assumption}

Let $g_j$ to represent the $j$-th n-gram in the candidate n-gram set $G$ and $q_i$ to represent the $i$-th query in the query set $Q$.
\lpms\ constructs a matrix $A$ of size $|Q| \times |G|$ and a vector $b$ of size $|Q|$ for constraint calculation, where $A_{i,j} = s_D(g_j) \cdot \mathbf{1}{g_j \in G(q_i)}$ and $b_i = \min_{g \in G(q_i)}{s_D(g)}$. This setup allows us to establish the constraint of the integer program, formalizing the {Assumption}\eat{two assumptions~\ref{assumption:lpms:1.1} and}~\ref{assumption:lpms:1} into the constraint $A x \geq b$.

However, the search space for all possible n-grams $G$ remains too large when the query set is large. \lpms\ adopts an iterative approach to select a {prefix-minimal} n-gram set from \free. In the $i$-th iteration, \lpms\ generates the candidate n-gram set $G_i$ with n-grams of size $i$ from all the {useless} n-grams from $G_{i-1}$. After solving the integer program in the $i$-th iteration, we insert the set of n-grams $I_i$ with $x_g = 1$ for all $g \in I_i$ into the index, and the remaining n-grams $G_i \setminus I_i$ are used to extend and generate $G_{i+1}$.

Solving the integer program is challenging, as the search space is $|G_i|^{O(|G_i|)}$~\cite{Lenstra_1983}. \lpms\ approximates the problem using linear programming with relaxation, as follows: 
\begin{align*}
    \text{minimize} \qquad & \sum_{g\in G} cv(g)x_g\\
    \text{subject to} \qquad&  A x \geq b; 0 \leq x_g \leq 1 \quad \forall g \in G 
\end{align*}
\subsubsection{Complexity Analysis}  
To calculate the space overhead, let's examine the sizes of each component. Both \textit{coverage} and the output of the linear program are vectors of size $|G_i|$. As previously discussed, $A$ has a size of $|Q| \cdot |G_i|$ and $b$ has a size of $|Q|$. Summing these, the space overhead for \lpms\ n-gram selection is $O(|Q| \cdot |G_i|)$.\eat{, which is $O(|Q| \cdot |\Sigma|^i)$. In practice, the size of $G_i$ for each iteration would be much smaller than $|\Sigma|^i$, the number of all possible n-grams of size $i$, since not all n-grams appear in the queries. Additionally, $G_i$ only contains n-grams whose prefixes are not already in the index $I$.  } During the algorithm runtime, \textit{coverage} is calculated using $s_D(g)$ and $s_Q(g)$. By utilizing \textit{coverage} for each n-gram rather than \textit{cover}, \lpms\ reduces the time complexity for constructing the \textit{coverage} vector to $O(|G_i| \cdot (|D| + |Q|))$ for each iteration. The linear program runs in polynomial time $O(|G_i|^{2.5})$~\cite{Vaidya_1989}. In practice, the size of the index key typically does not exceed 10, a number also used as the upper bound for n-gram size in \free. Therefore, we can consider the small number of iterations as a constant, making the overall computational complexity of \lpms\ $O(\sum_{i \in [10]}\{|G_i|^{2.5} + |G_i| \cdot (|D| + |Q|)\})$.

\subsection{VGGraph}\label{subsec:method:vggraph}

{The n-gram selection strategy from \vgg~\cite{Qiu_2025} combines the insights from \free\ and \best\ to select the n-grams. }
\subsubsection{Selection Strategy}
{The idea is to select the most cost-effective n-gram from a set $S$ to cover characters in an uncovered set $E$. Each n-gram is evaluated based on a cost-efficiency ratio (a notion similar to \textit{cover} from~\Cref{def:best:cover}), where the cost is size of the inverted index for the n-gram and the efficiency is the number of characters it helps cover in $E$. \vgg\ approximate a perfect set cover with efficiency by adopting \free's idea of \textit{Usefulness}, and only extend the useful n-grams. By maintaining the uncovered set $E$ that is continuously updated as the n-grams are picked, \vgg\ ensures that prefix-free n-grams are selected preferentially.}

\subsubsection{Complexity Analysis}
{\vgg's index key selection consists of two steps: (1) recursive extension of grams exceeding the frequency threshold $c$, and (2) greedy n-gram selection via the set-cover approximation. Given a dataset $D$, the complexity of recursively extending grams is $O(|D|\cdot q_{\min})$, as each character position is indexed exactly once and each gram is extended at most once. The greedy set-cover step selects, for each regex, a minimal subset of grams from the candidate set $G(Q)$. For a regex with literal size $n$, the greedy selection has a complexity of $O(|G(Q)|\cdot n)$ per query. The total runtime complexity for key selection, combining both steps, is thus $O(|D|\cdot q_{\min} + |Q|\cdot|G(Q)|\cdot n)$. The corresponding runtime space complexity during key selection is $O(|D|+|G(Q)|)$, dominated by storing candidate grams and their counts.}

\section{Experiment Setup}
Due to the absence of source code from the papers, we implemented the three n-gram selection methods ourselves. 
{For a fair comparison of the n-gram selection methods, we use the same index structure (inverted index) and same basic query matcher for all experiments. We include the implementations of other index structures and query plans per the original papers in our codebase for reference. }
\eat{For each method, we also developed their corresponding query parsers (if any) and index structures. }\eat{The index building and lookup methods are integrated into a single framework that also handles workload loading, configuration settings, and result generation.}

\subsection{Benchmark Framework}
\begin{figure}[!tp]
\centering
  \centering
  \includegraphics[width=0.8\linewidth, scale=0.8]{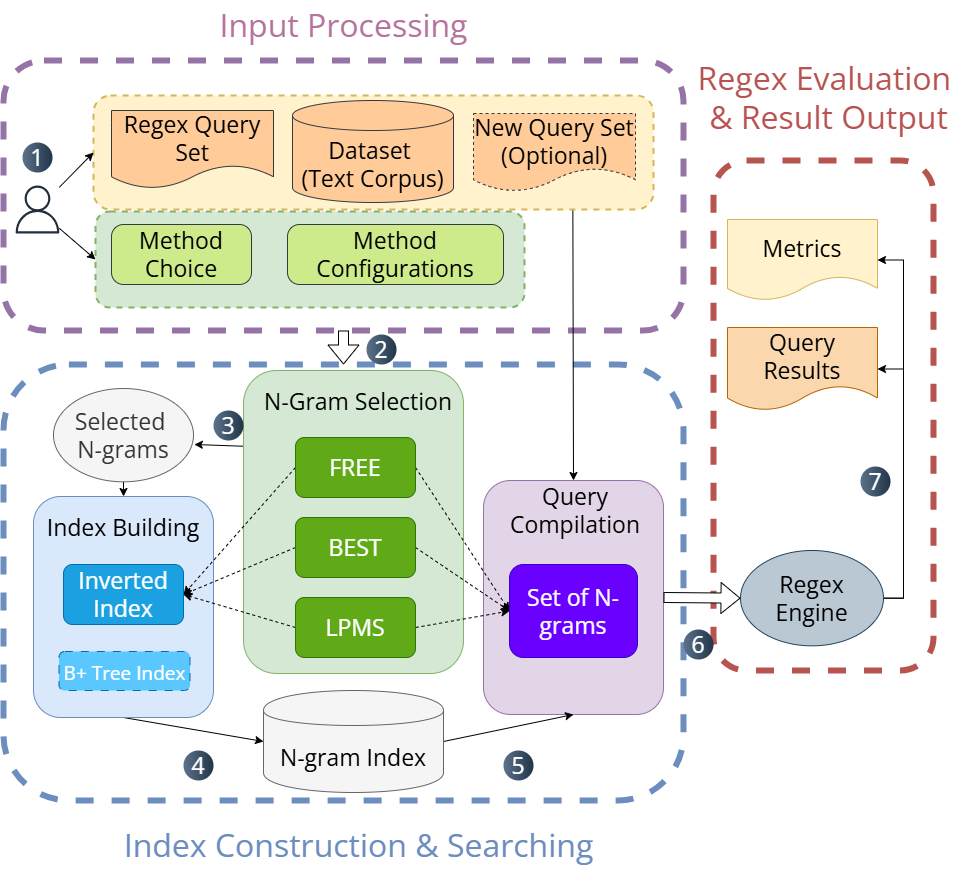} 
  \caption{Benchmarking framework overview.}
  \label{fig:framework}
  \vspace{-1em}
\end{figure}

The benchmark framework is designed to facilitate the comparison of n-gram selection techniques (FREE, BEST, and LPMS) for regex indexing, with a focus on modularity, extensibility, and reproducibility across a range of workloads. We summarize its detailed architecture and workflow in~\Cref{fig:framework}.

\looseness=-1 The end-to-end process follows a seven-step sequence, as illustrated in the framework diagram~\Cref{fig:framework}. t begins with user-specified inputs (dataset, regex queries, method, and configurations), which are processed by the framework. The selected n-gram strategy analyzes the workload to identify optimal n-grams for index construction. These n-grams are then used to build an index (e.g., inverted index or B+-tree). Next, the index search plan extracts n-grams from regex literals to guide candidate filtering. The regex engine verifies these candidates, discards false positives, and computes performance metrics. Finally, results are aggregated and exported. The pipeline is organized into three phases—input processing, index construction, and regex evaluation—with components grouped accordingly in~\Cref{fig:framework}. \eat{The framework supports diverse configurations and workloads while enabling direct comparisons of n-gram selection strategies. The framework's modular design allows for changes to the regex engine, index structure, or index search plan without altering other components.}\eat{ \sd{Let's label each phase in the diagram too.}}

\introparagraph{Input Processing} The framework begins by accepting a text file containing the string dataset and a text file with the regex query set. Users also specify n-gram selection methods and their parameters, including n-gram length, selectivity thresholds, maximum number of n-grams, thread counts, and more for index building. Optionally, a new query set can be provided at runtime, enabling dynamic evaluation of the index against unseen query workload. These inputs are standardized to a unified format across experiments\eat{, as summarized in the purple dashed bounding box in~\Cref{fig:framework}}. During this phase, all data in the workload is loaded into memory.

\introparagraph{Index Construction and Searching}
\eat{\TODO{Should we remove the B+-tree and query plan generation from the section and the overview figure all together, or just state that we are using inverted index and simple index lookup strategy (set of n-grams query)} \sd{Yes, I think we remove the B-tree and just use inverted index throughout for a fair comparison.}}
In the index construction phase, one of the three n-gram selection strategies is applied to the workload. The selected n-grams are used to build an index.\eat{, such as an inverted index or a B+-tree.} This phase leverages multi-threaded execution if specified by the user. After the index is built, the index search plan is compiled if necessary. 
\eat{{User can select to build either inverted index (used in the original paper of \free\ and \lpms) or B+-tree (used in the original paper of \best). }}
{An inverted index with data string IDs is built with the selected n-grams.
In the simple query plan, posting lists corresponding to the set of n-grams that exists in the regex query literals are unioned together later for full regex evaluation.} 
\eat{Only \free\ requires an additional intermediate step to generate an index search tree. 
For a fair comparison, we use inverted index after all n-gram selection methods and simple query 
 plan for index lookup.}
\eat{Each strategy has its own corresponding type of index structure and index search plan, as shown in the blue dashed bounding box in~\Cref{fig:framework}. \free\ and \lpms\ use inverted indexes, while \best\ uses a B+-tree for the index structure. During index lookup, all n-grams in both the set of regex query literals and index keys are extracted. Only \free\ requires an additional intermediate step to generate an index search tree. }

\introparagraph{Regex Evaluation and Result Output}
\looseness=-1 The final phase processes all possible data points after index lookup and validates them using a regex engine (e.g., RE2, PCRE2) to perform exact matches and eliminate false positives. Metrics such as index construction time, runtime memory consumption, workload processing time, and false-positive rates are measured during index construction or after regex evaluation.

\begin{table}[t]
\footnotesize
\centering
\captionof{table}{Workload $W=(Q,D, \Sigma)$ statistics. We use $\bar{|d|}$ to denote average data record size (in bytes) and $\bar{TP}$ to denote the average number of actual matches of all queries. \wlsyn\ has two query sets -- the first set of 500 queries that is used to build the index, and a second set of 100 queries that is used to test the impact of queries that are not indexed.}
\begin{tabularx}{\columnwidth}{ L| r R  r  r  r  R}
  \toprule
 Workload & $|Q|$ & $|D|$ & $|\sum|$ & $\bar{|d|}$ & $\bar{TP}$ & Dataset Size \\ \midrule
\wlweb & 10 & 695,565 & 255 & 68,650 & 86,524 & 47 \scriptsize{GB}\\
\wldblp & 1000 & 305,798 & 122 & 38 & {914} & 32 \scriptsize{MB}\\ 
\wlprotein & 101 & 111,788 & 22 & 416 & 722 & 45 \scriptsize{MB} \\
\wltraffic & 4 & 2,845,343 & 99 & 405 & 92,042 & 1.1 \scriptsize{GB} \\
\wlsql & 132  & 101,876,733 & 114 & 139 & 50,356 & 14 \scriptsize{GB}\\
\wlenron & {107} & \makecell[r]{{ 517,401}} & {120} & {2719} & \makecell[r]{{54,242}} & \makecell[r]{{2.6 \scriptsize{GB}}}\\
\wlsysy & {18} & \makecell[r]{{890,623,051}} & {118} & {101} & \makecell[r]{{4,653,057}} & \makecell[r]{{100 GB}} \\
\wlsynnew & \makecell[r]{ {vary}} & \makecell[r]{{50,000}} & {16} & {200} & {vary} & \makecell[r]{{9.6 \scriptsize{MB}}}\\
\bottomrule 
\end{tabularx}
\label{tab:wl_summary}
\end{table}

\subsection{Workloads}

We use several real-world text datasets and queries with varying numbers of queries, data strings, alphabet sizes, average string lengths, and average matches per query for a comprehensive analysis. We summarize the workload characteristics in~\Cref{tab:wl_summary}. Among the workloads, three (\wlweb, \wldblp, and \wlprotein) are those used by the original \free, \best, and \lpms\ papers to evaluate their methods. The other four workloads (\wltraffic, \wlsql, \wlenron, \wlsysy) are more recent and feature larger datasets.

\introparagraph{\wlweb} This workload was used by \free. In the original paper, the authors utilized 700,000 random web page HTML files downloaded in 1999, along with 10 regex queries suggested by researchers at IBM Almaden~\cite{FREE}. While the exact query set is included in the paper, the dataset is not. Since we could not locate the original dataset, we constructed a similar dataset using web pages from 2013 stored in Common Crawl~\cite{common_crawl}. We chose 2013 data because it is relatively close to 1999, ensuring that most of the regexes constructed for the 1999 dataset would still have matches in the 2013 dataset. We selected the web pages to ensure a relatively balanced number of matches for each regex query.

\introparagraph{\wldblp} This workload was used by \best. The authors collected 305,798 (Author-Name, Title-of-Publication) tuples as the dataset. We used the DBLP-Citation-network dataset~\cite{Tang_2008} and  selected the same number of entries uniformly at random. The query set is constructed by choosing author last names from the pool of author names uniformly at random to obtain $1000$ queries. We then constructed the regex query for each last name by appending \verb|.+| and a space in front of the last name.

\introparagraph{\wlprotein} This workload was used by \lpms. Following the paper's description, we selected 100,000 protein sequences from the PFAM-A database~\cite{Mistry_2020} and chose 100 Prosite signatures~\cite{Sigrist_2012}, transforming them into 100 regular expressions.

\introparagraph{\wltraffic} This is an open-source dataset containing descriptions of traffic accidents in the United States from February 2016 to March 2019~\cite{Moosavi19}. The dataset has 2,845,343 strings. The authors included four regex queries on the dataset in the original table.

\introparagraph{\wlsql} This is a production workload consisting of in total 101,876,733 log messages generated by Microsoft SQL Server and 132 regex queries used for data analysis tasks on this dataset.

\introparagraph{\wlenron}
{This is a production workload consists of more than 50 thousand of real emails and 107 regex queries. The dataset is the May 7, 2015 Version of Enron dataset~\cite{enron}, and the queries are from Avatar search engine~\cite{avatar} collected by Chen et al.~\cite{chen_2012}}.

\introparagraph{\wlsysy}
{This is a real production workload consists of 890 millions of logs from Azure Data Explorer (ADX) and 18 regexes for an log analysis task on this dataset by an data scientist. }

\introparagraph{\wlsynnew}
{This workload is based on the method described in the paper proposing \lpms~\cite{FAST}. The workload consists of 50,000 data records with a fixed alphabet size of 16. We varied the query set size across $[500, 1000, 2000, 5000]$ and explored different levels of query selectivity ranging from 0.001 to 0.5.\eat{in $[0.001, 0.005, 0.01, 0.02, 0.05, 0.1,$
$ 0.2, 0.5]$. }
}
\subsection{Metrics}\label{subsec:metrics}

We compare the three methods on the following aspects:

\eat{\paragraph{N-Gram selection time ($T_I$)} }

\introparagraph{N-Gram Index Construction Time ($T_I$)} We measure the time required to select all n-grams for indexing and building the index, denoted by $T_I$, after loading the necessary data into memory. This aspect is crucial, as we deal with much larger datasets than when these n-gram selection algorithms were originally presented. The selection time may become prohibitively long, rendering the methods impractical. \eat{Since \best\ and \lpms\ populate the index while making selections, for a fair comparison, we record the total time to select n-grams and construct the index, denoted by $T_I$.}

\introparagraph{Precision} We use micro-average precision to compare the filtering power of an index. This metric best describes the effectiveness of the selected n-grams and is commonly used in information retrieval~\cite{Zhai_2001, Liu_2024}\eat{ \sd{add citations}}. It provides a balanced measure across all queries with different numbers of matches, without giving disproportionate weight to any individual query. It aggregates actual data records that match the regex query (true positives, denoted by $TP$) and data records that pass the index filtering but do not match the regex query (false positives, denoted by $FP$). For a workload $W = (Q,D)$, the overall precision on index $I$ is:
$$Prec_W = \frac{\sum_{q\in Q} \#TP_q}{\sum_{q\in Q} \#TP_q + \sum_{q\in Q} \#FP_q} $$

\introparagraph{Query Time ($T_Q$)} We measure the time to run the workload, denoted by $T_Q$, to demonstrate the runtime gains provided by the index. 

\introparagraph{Runtime Space Usage ($S_Q$)} This metric is the peak space used when running a experiment.

\introparagraph{Index Size ($S_I$)} We measure the size of the index constructed, which is the total size of the n-gram keys, the posting lists, and necessary index data structure (such as the inverted index).
\eat{ \jmp{What about the B+-tree when it is built?} \Ling{Ling: We store the list of data strings as value for B+tree as well.}
}

\subsection{Hardware and Implementation Details} \label{subsec:implementation}
\looseness=-1 We implemented n-gram selection methods \free, \best, \lpms, {\vgg}, and {\trigram} with the index construction and matching methods mentioned in the benchmarking framework. We implemented \free' query parser and \best\ with a B+-tree in the codebase as well. We used Gurobi Optimizer~\cite{gurobi} as the linear program solver. All experiments were conducted on an Azure Standard\_E32-16ds\_v5 machine with an Intel(R) Xeon(R) Platinum 8370C CPU @ 2.80GHz, 16 vCPUs, 256 GB of memory, and 1 TB of disk storage. Our benchmarking framework is written in C++17 and compiled with the -O3 flag. Google's RE2 (release version 2022-06-01)~\cite{re2}\eat{\sd{add this as a citation}} as the regex engine for {regex evaluation}\eat{verification \sd{maybe we should call verification as just evaluation throughout.}}.

\looseness-1 Each method has its own configurable parameters. For \best\ and \free, we vary the selectivity threshold $c$ between 0.01 and 0.7. \free\ also varies the maximum n-gram length ($\max_n \in \{2, 4, 6, 8, 10\}$) and whether the selected n-grams are prefix-minimal or pre-suf-minimal. All experiments are capped at 3 hours; if \best\ exceeds this, we reduce the workload to 0.1–0.85\%.
\lpms\ supports both deterministic (\lpms-D) and randomized (\lpms-R) relaxations. We also cap the number of n-grams for \lpms\ if runtime is excessive. All methods run with 16 threads. To ensure fair comparison, we match index sizes across methods. Since \free\ and \lpms\ don’t natively support key limits, we implement early stopping via a max-key parameter for all three methods. By providing an optional parameter of max number of keys\eat{ ($\max_n$)}, the n-gram selection methods will stop once the limit is reached. 
{Each configuration of a selection method generates a specific set of n-grams of varying sizes. To ensure a fair comparison, we compare selected sets of similar sizes, as indexing all possible n-grams would yield the highest precision for the workload. Since each methods has different configurations, and those configuration values interfere with each other in a non-linear way, isolating one configuration and study its behavior as value changes is not meaningful. For each n-gram selection method under a specific key number constraint $K$ $s.t.$ $|I| \leq K$, we select the configuration with the highest precision. }
\eat{Note that throughout the section, we report the numbers for space usage and query performance for the configuration that achieves the best possible precision given a constraint on $K$.}
We choose the values of max number of keys{, $K$, } on a case-by-case basis for each workload, as specified in~\Cref{sec:expr}.
 
\section{Experiments} \label{sec:expr}

In this section, we run the aforementioned workloads on {the indexes build from different n-gram selection methods} with varying parameters. In particular, we aim to answer the following questions:

\begin{enumerate}[label=\textbf{Q.\arabic*}, ref=Q.\arabic*]
    \item What is the usefulness of each n-gram selection method? We evaluate this through precision and $T_Q$.\label{q:performance}
    \item What is the index construction overhead for each method? We evaluate this through $T_I$, $S_Q$, and $S_I$. \label{q:overhead}
    \item How do the characteristics of each workload affect the usefulness of n-gram selection methods? \label{q:workloads}
\end{enumerate}

We answer question~\ref{q:performance} in~\Cref{subsec:expr:query_perf} and question~\ref{q:overhead} in~\Cref{subsec:expr:index_overhead}. 
To answer question~\ref{q:workloads}, we discuss the impact of characteristics of different real-world workloads on index construction overhead in~\Cref{subsec:expr:index_overhead} and on index effectiveness in~\Cref{subsec:expr:query_perf}. We also use present a more thorough ablation study on impact of workload parameters such as the query set size $|Q|$, dataset size $|D|$, , query selectivity, etc. in~\Cref{subsec:expr:workload_parameters}.

\eat{
\renewcommand{\arraystretch}{0.8} 
\begin{table}[t]
\centering
\captionof{table}{Index cost and query performance on \wldblp{}. }
\scalebox{1}{
\begin{tabularx}{\columnwidth}{ C | C | R  R  R  R |  R }
\toprule
 $K$ & Method & $T_{Q}$ s & $T_{I}$ s &$S_{Q}$\scriptsize{GB} & $S_{I}$\scriptsize{MB} & $Prec$  \\ \midrule
\multirow{3}{*}{\bf{150}}& \best  & \cellcolor{green!50}{19.7} & 533 & 2.194 & \cellcolor{green!50}{33.433} & \cellcolor{yellow!50}{0.235} \\
& \free  & 25.3 & \cellcolor{green!50}{2} & 0.171 & 37.299 & 0.051 \\
& \lpms  & 19.8 & 3 & \cellcolor{green!50}{0.155}& 34.028 & 0.048 \\ \midrule
\multirow{5}{*}{\bf{500}}& \best  & \cellcolor{green!50}{15.2} & 8762 & 2.670 & 54.306 & \cellcolor{yellow!50}{0.242} \\
& \free  & 20.6 & {2} & 0.242 & 66.774 & 0.165 \\
& \lpms  & 17.7 & 6 & \cellcolor{green!50}{0.158} & {40.657} & 0.070 \\ 
& \trigram  & 54.0 & \cellcolor{green!50}{0} & 0.208 & \cellcolor{green!50}{14.153} & 0.006 \\
& \vgg  & 22.1 & 0 & 0.276 & 14.198 & 0.035 \\ \midrule
\multirow{5}{*}{\bf{1000}}& \best  & \cellcolor{green!50}{15.2} & 8762 & 2.670 & 54.306 & \cellcolor{yellow!50}{0.242} \\
& \free  & 16.5 & 1 & \cellcolor{green!50}{0.287} & 75.354 & 0.219 \\
& \lpms  & 17.0 & 808 & 1.309 & 41.499 & 0.071 \\ 
& \trigram  & 40.8 & \cellcolor{green!50}{0} & \cellcolor{green!50}{0.194} & \cellcolor{green!50}{20.405} & 0.009 \\
& \vgg  & 18.8 & 0 & 0.275 & 22.076 & 0.060 \\ \midrule
\multirow{5}{*}{\bf{2000}}& \best  & \cellcolor{green!50}{15.2} & 8762 & 2.670 & 54.306 & 0.242 \\
& \free  & 15.9 & 1 & \cellcolor{green!50}{0.264} & 68.442 & \cellcolor{yellow!50}{0.321} \\
& \lpms  & 17.0 & 808 & 1.309 & {41.499} & 0.071 \\ 
& \trigram  & 27.2 & \cellcolor{green!50}{0} & 0.277 & \cellcolor{green!50}{31.874} & 0.017 \\
& \vgg  & 15.5 & 0 & 0.282 & 39.136 & 0.167 \\ \midrule
\multirow{5}{*}{\bf{3000}}& \best  & 15.2 & 8762 & 2.670 & 54.306 & 0.242 \\
& \free  & {14.9}  & {2} & \cellcolor{green!50}{0.304} & 83.069 & {0.524} \\
& \lpms & 17.0 & 808 & 1.309 & {41.499} & 0.071 \\ 
& \trigram  & 27.2 & \cellcolor{green!50}{0} & 0.277 & \cellcolor{green!50}{31.874} & 0.017 \\
& \vgg  & \cellcolor{green!50}{14.8} & 0 & 0.284 & 79.833 & \cellcolor{yellow!50}{0.709} \\ \midrule
\multirow{5}{*}{\bf{4000}}& \best  & 15.2 & 8762 & 2.670 & 54.306 & 0.242 \\
& \free  & {14.5} & {2} & \cellcolor{green!50}{0.304} & 83.069 & {0.524} \\
& \lpms  & 17.0 & 808 & 1.309 & {41.499} & 0.071 \\ 
& \trigram  & 27.2 & \cellcolor{green!50}{0} & 0.277 & \cellcolor{green!50}{31.874} & 0.017 \\
& \vgg  & \cellcolor{green!50}{14.4} & 3 & 0.310 & 83.639 & \cellcolor{yellow!50}{0.725} \\
\bottomrule
\end{tabularx}}
\label{tab:dblp1k_summary}
\end{table}
}

\begin{figure}[!tp]
\centering
\begin{subfigure}{0.7\columnwidth}
  \centering
  \includegraphics[width=0.95\linewidth, scale=0.95]{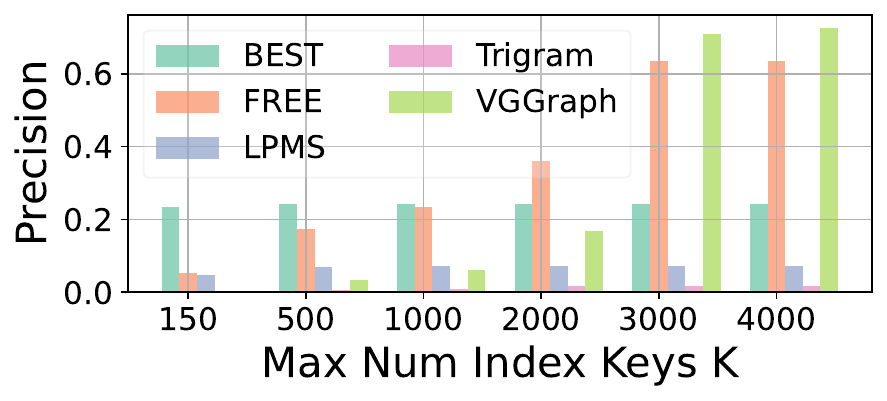}  
  \vspace{-1em}
  \caption{Precision}
  \label{fig:dblp_prec}
\end{subfigure}\\
\begin{subfigure}{0.7\columnwidth}
  \centering
  \includegraphics[width=.95\linewidth, scale=0.95]{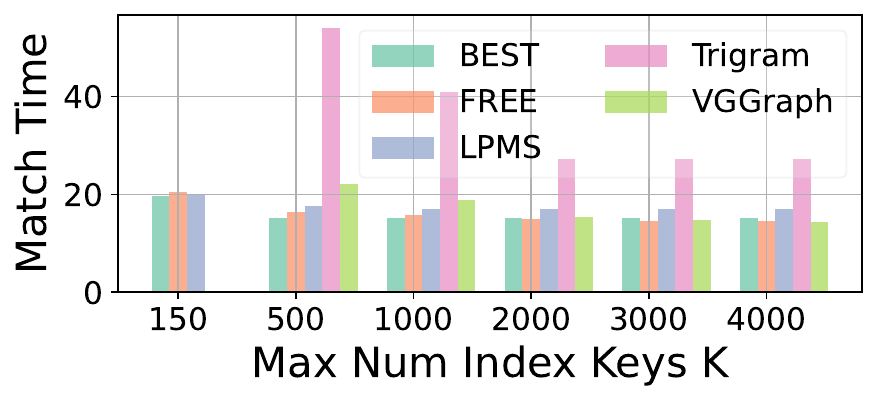}  
  \vspace{-1em}
  \caption{Query Time $T_Q$ (s)}
  \label{fig:dblp_queryTime}
\end{subfigure}
  \vspace{-1em}
\caption{\wldblp.\eat{ We use the bar plot to show the variable value of the highest-precision index for each method under different key number constraints $K$.}}
\label{fig:dblp1000_perf}
\vspace{-1.5em}
\end{figure}

\subsection{Query Performance} \label{subsec:expr:query_perf}
To address question~\ref{q:performance}, we compare the performance gains from using indices built with n-grams selected by the three methods, focusing on filtering precision. Higher precision indicates a more effective n-gram set. We also measure the regex query runtime $T_Q$, which reflects precision since the index structure and matching strategy remain constant.

Generally, indexing more n-grams improves precision. However, each method behaves differently. \best\ often starts with the highest precision when the key budget $K$ is small, but its precision gains plateau due to time constraints (e.g., 3-hour index build). In contrast, \free\ begins with lower precision—especially for small query sets—but improves significantly as $K$ increases, often achieving the highest precision overall.
{Precision of \vgg\ is low when number of keys is small, but quickly rises as number of keys gets larger.}
\lpms\ has similar precision as \best\ when the numbers of key is small and usually grows insignificantly as the number of n-grams selected increases. {\trigram\ typically has very low precision when the number of index keys fixed is low.} We now investigate the methods' precision for different workloads with varying characteristics. 

\subsubsection{Workload \wldblp}
We evaluate \best, \free, \lpms, and \vgg\ across various input configurations, and applying the same key limit to \trigram. Performance is compared based on index size, focusing on configurations with at most 5000 n-grams.

As shown in~\Cref{fig:dblp_prec}, \best($c=0.5$) achieves a precision of 0.235 for $K \leq 150$, outperforming \free($\max_n=2$, $c=0.5$) and \lpms-D, which yield much lower precision (0.051 and 0.048, respectively) at a similar index size.  At $K \leq 500$, the 363 n-grams selected by \best($c=0.5$) achieves significantly better precision compared to \lpms-D. With similar index construction overhead, \free($\max_n=2$,$c=0.5$) achieves higher precision than \lpms-D. 
\lpms\ tends to select infrequent n-grams, which may not benefit overall workload filtering. Although \best\ improves slightly with more keys (e.g., +0.007 precision and +4.5s query performance from $K=150$ to $K=500$), gains are modest. {Meanwhile, \vgg($\max_n=2$,$c=0.5$) builds an index with 215 n-grams but with a low precision of 0.035. }

When relaxing the index size constraint to 1000 n-grams, \free\ ($\max_n=4$,$c=0.5$) generated an index of size 810 that achieved a precision of 0.219, slightly lower than \best($c=0.5$). 
{\vgg ($max_n = 3$, $c=0.5$)'s precision increase is insignificant.}
As the number of n-grams allowed in the index increases, starting from 2000, more indices from different configurations of \free\ fall into this range. The best configurations of \free($\max_n=2$,$c=0.2$) achieve higher precision than \best. The precision increases from 0.321 at 2000 keys to 0.524 at 4000 keys upper limit by \free($\max_n=4$,$c=0.15$). {Precision of \vgg\ increases significantly from 0.167 to 0.725 as number of keys increases from 1014 to 3681.}

\wldblp{} runs much faster with all index configurations than the baseline (i.e. using no indexes at all) of 99 seconds. {Even when \trigram\ indexes are built using only the first few trigrams—yielding low precision between 0.006 and 0.017, they still achieves performance improvement of $1.8\times$ to $3.6\times$.} Looking at~\Cref{fig:dblp_queryTime}, however, the improvement in workload performance is significant compared to the change in precision. The precision of \free\ increases around 10 times as $K$ increases from 150 to 3000; its query time decreases by 41\%.
When $K=150$, the precision of \best\ is significantly larger than \free\ and \lpms\ in~\Cref{fig:dblp_prec}, the performance difference for \best\ and \lpms\ is not significantly different.
\eat{On exception is when $K$ increases from 150 to 500, and \best\ gets a precision raise of 30\%,  }
In the \wldblp\ workload, each string in the dataset is short, resulting in very quick regex evaluation, and therefore the query time reduction brought by precision increase is less significant. {\trigram's precision increases as number of keys increases. When indexing all possible 28,275 trigrams, we can get a high precision of 0.97. However, the query time is 14.7s, similar to \best\ with less than 500 keys. This suggests that indexing all trigrams is not cost-efficient in terms of storage.}

\insight{\free\ performs best on the \wldblp\ workload. Generally, if the workload has a large number of queries that are not skewed (i.e., all very similar and covering a small subset of the dataset) and the strings in the dataset are not long, it is best to choose \free. {\vgg\ significantly improve its precision only when the number of keys increase.} When the query set is large, \best\ and \lpms\ require significant computation time and memory, owing to their higher time and space complexities. Additionally, for a large-sized and balanced query set, frequent n-grams in the datasets are likely to be covered in the queries as well. }

\eat{\begin{figure}[!tp]
\centering
\begin{subfigure}{0.45\columnwidth}
  \centering
  \includegraphics[width=0.95\linewidth]{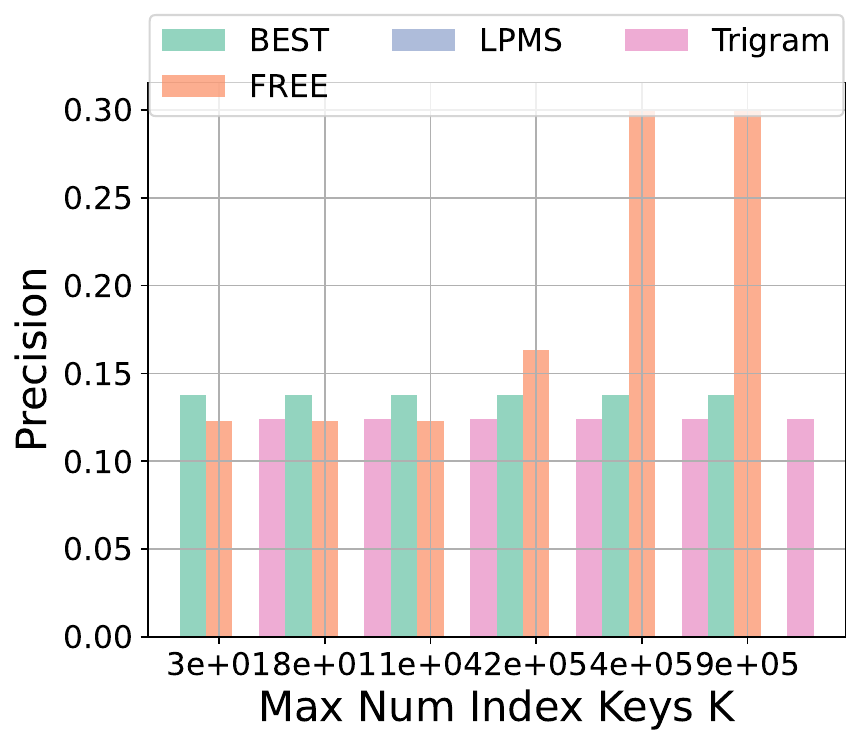}  
  \caption{Precision}
  \label{fig:web_perc}
\end{subfigure}
\begin{subfigure}{0.45\columnwidth}
  \centering
  \includegraphics[width=.95\linewidth]{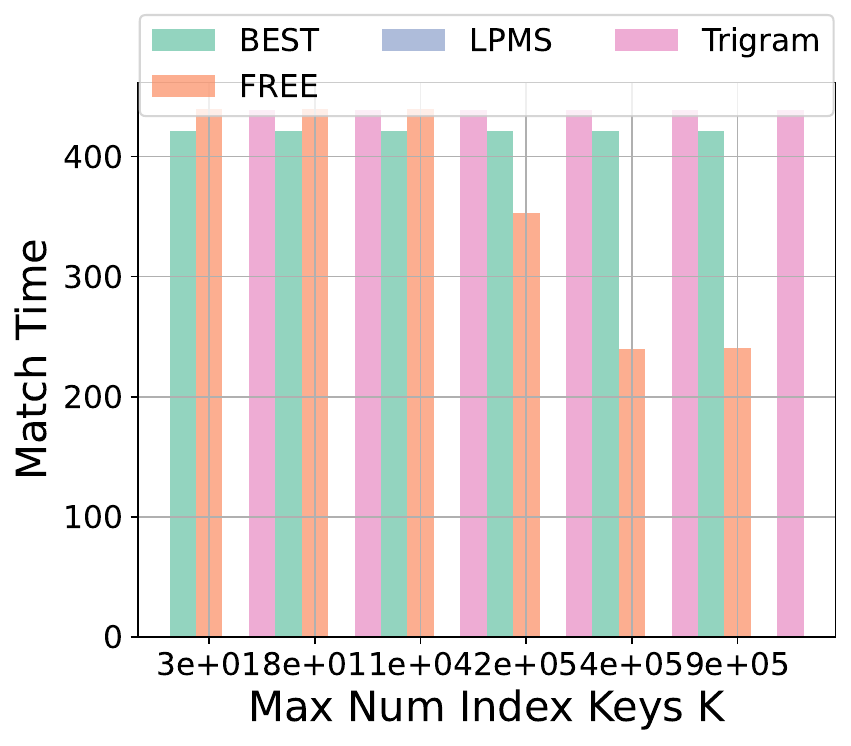}  
  \caption{Query Time $T_Q$ (s)}
  \label{fig:web_queryTime}
\end{subfigure}
\caption{\wlweb.\eat{ We use the bar plot to show the variable value of the highest-precision index for each method under different key number constraints $K$.}}
\label{fig:web_perf}
\end{figure}
}

\begin{table}[t]
\centering
\captionof{table}{Index cost and query performance on \wlweb.}
\vspace*{-1em}
\scalebox{0.9}{
\begin{tabularx}{\columnwidth}{ C | C | R  R  R  R |  R }
\toprule
 $K$ & Method & $T_{Q}$ s & $T_{I}$ s &$S_{Q}$ \scriptsize{GB} & $S_{I}$ \scriptsize{MB} & $Prec$  \\ \midrule
\multirow{2}{*}{\bf{5}}& \best  & \cellcolor{green!50}{422} & 8587 & 150 & 0.1 & \cellcolor{yellow!50}{0.138} \\ 
& \free  & 440 & \cellcolor{green!50}{382} & \cellcolor{green!50}{101} & \cellcolor{green!50}{0.1} & 0.123  \\ 
\midrule
\multirow{2}{*}{$\mathbf{1.7\times 10^5}$}& \best  & 422 & 8587 & \cellcolor{green!50}{150} & \cellcolor{green!50}{0.1} & 0.138\\ 
& \free  & \cellcolor{green!50}{353} & \cellcolor{green!50}{1184} & 159 & 357.9 & \cellcolor{yellow!50}{0.163} \\
\midrule
\multirow{2}{*}{$\mathbf{3.6\times 10^5}$}& \best  & 422 & 8587 & \cellcolor{green!50}{150} & \cellcolor{green!50}{0.1} & 0.138  \\ 
& \free  & \cellcolor{green!50}{240} & \cellcolor{green!50}{5965} & 240 & 2007.4 & \cellcolor{yellow!50}{0.301} \\
\bottomrule
\eat{\midrule
\multirow{2}{*}{$\mathbf{1.5\times 10^6}$}& \best  & 422 & 8587 & \cellcolor{green!50}{150} & \cellcolor{green!50}{0.1} & 0.138 \\ 
& \free  & \cellcolor{green!50}{239} & \cellcolor{green!50}{7480} & 233 & 524.9 & \cellcolor{yellow!50}{0.302} \\ \bottomrule}
\end{tabularx}}
\label{tab:web_summary}
\vspace{-1em}
\end{table}

\vspace{-1em}
\subsubsection{Workload \wlweb{}} 
We evaluate \wldblp, a workload with significantly fewer queries than \wldblp, using HTML files from \wlweb, which has the longest average string length among all workloads. Results are summarized in~\Cref{tab:web_summary}, with a baseline (no index) query time of 437s.

At $K=5$, \best($c=0.1$) selects just four n-grams and achieves 0.138 precision. A manually early-stopped \free($\max_n=2$, $c=0.02$) with five n-grams yields a similar 0.123 precision. To match that precision without early stopping, \free($\max_n=2$, $c=0.5$) requires $1.7 \times 10^5$ keys. This difference stems from selection strategies: \best\ chooses n-grams from the query–dataset intersection, limiting candidates when queries are few. In contrast, \free\ selects from the dataset alone, resulting in a much larger candidate pool. Despite requiring more keys, \free\ achieves 0.025 higher precision than \best\ and delivers a 1.2× speedup in query time—reducing it from 421.6s to 353.0s as shown in~\Cref{tab:web_summary}.

When the number of keys allowed $K$ increases from $1.7\times 10^5$ to $3.6\times 10^5$ , the precision of the index by \free($\max_n=2$,$c=0.7$) increases significantly from 0.163 to 0.301.
At $K = 3.6 \times 10^5$, \free\ delivers a query time that is 1.47× faster than its own earlier version and 1.75× faster than \best. Compared to the overall precision and query time trend for \wldblp\ workload in~\Cref{fig:dblp1000_perf}, the precision increase of \free\ in \wlweb\ corresponds to a more significant workload runtime decrease from 353 seconds to 240 seconds. Since each string in \wlweb\ is a long HTML page, filtering more strings upfront yields greater performance benefits. While \free’s larger index may not always justify its size relative to \best, it offers better robustness to unseen queries on the same dataset. Due to lack of space, we defer the reader to \robustTest\ of the full paper for detailed experiments. {\vgg\ and }\lpms\ did not finish for this dataset within the set time frame due to the large average length per string of the dataset.

\insight{\best\ is suitable for a workload like \wlweb\ where the query set is small. \best\ selects the set of n-grams that achieves near-optimal precision, with the optimality resulting from the long computation time. With a small query set, the index construction time is reasonable. However, for a workload where each document entry is large, the precision and robustness of the index become more important, especially if the dataset is likely to be queried with different regex queries.
}

\subsubsection{Workload \wlprotein}
\wlprotein\ has the largest query size to number of records in the dataset ratio among the real-world workloads. The strings in the dataset have a  mean length of 416 characters. Two distinguishing characteristics of \wlprotein\ are: 1) it has the smallest alphabet size of 22, and 2) it has very short literal components in its the regex queries.
{The short literal components gives little opportunity for n-gram index improvement, and all methods created indices with low precision and performance improvement.}
In fact, the baseline workload runtime is 154 seconds. 

\eat{\begin{table}[t]
\centering
\captionof{table}{Index cost and query performance on \wlprotein.\eat{ For each n-gram selection method under a specific key number constraint $K$ $s.t.$ $|I| \leq K$, we select the configuration with the highest precision. }}
\vspace*{-1em}
\scalebox{1}{
\begin{tabularx}{\columnwidth}{ C | C | R  R  R  R |  R }
\toprule
 $K$ & Method & $T_{Q}$ s & $T_{I}$ s &$S_{Q}$\scriptsize{GB} & $S_{I}$\scriptsize{MB} & $Prec$  \\ \midrule
\multirow{3}{*}{\bf{50}}& \best  & \cellcolor{green!50}{141.5} & 400 & 0.931 & 9.4 & \cellcolor{yellow!50}{0.00826} \\
& \free  & 154.0 & \cellcolor{green!50}{2} & \cellcolor{green!50}{0.227} & \cellcolor{green!50}{1.5} & 0.00651 \\
& \lpms  & 151.8 & 14 & 0.263 & 26.2 & 0.00708 \\ \midrule
\multirow{3}{*}{\bf{100}}& \best  & \cellcolor{green!50}{139.7} & 856 & 0.967 & 14.9 & \cellcolor{yellow!50}{0.0089} \\
& \free  & 153.6 & \cellcolor{green!50}{2} & \cellcolor{green!50}{0.242} & \cellcolor{green!50}{5.7} & 0.00652 \\
& \lpms  & 151.8 & 14 & 0.263 & 26.2 & 0.00708 \\ \midrule
\multirow{3}{*}{\bf{500}}& \best  & \cellcolor{green!50}{139.7} & 856 & 0.967 & \cellcolor{green!50}{14.9} & \cellcolor{yellow!50}{0.00890} \\
& \free  & 150.8 & \cellcolor{green!50}{3} & 0.595 & 141.4 & 0.00687 \\
& \lpms  & 151.8 & 14 & \cellcolor{green!50}{0.263} & 26.2 & 0.00708 \\  \bottomrule
\end{tabularx}}
\label{tab:protein_summary}
\end{table}
\Cref{tab:protein_summary} shows the results. }
\eat{\TODO{add figures}}

\begin{figure}[!tp]
\centering
\begin{subfigure}{0.42\columnwidth}
  \centering
  \includegraphics[width=0.95\linewidth]{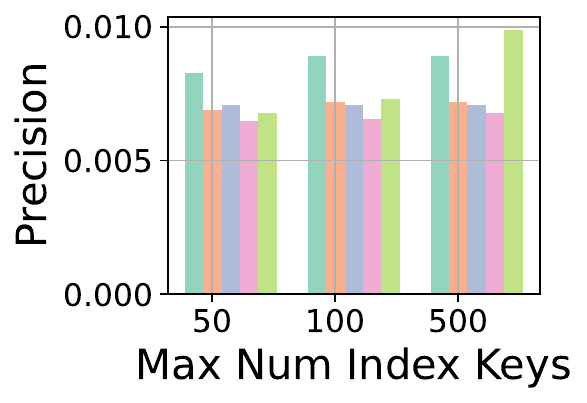}  
    \vspace{-0.5em}
  \caption{Precision}
  \vspace{-1em}
  \label{fig:protein_perc}
\end{subfigure}
\begin{subfigure}{0.57\columnwidth}
  \centering
  \includegraphics[width=.95\linewidth]{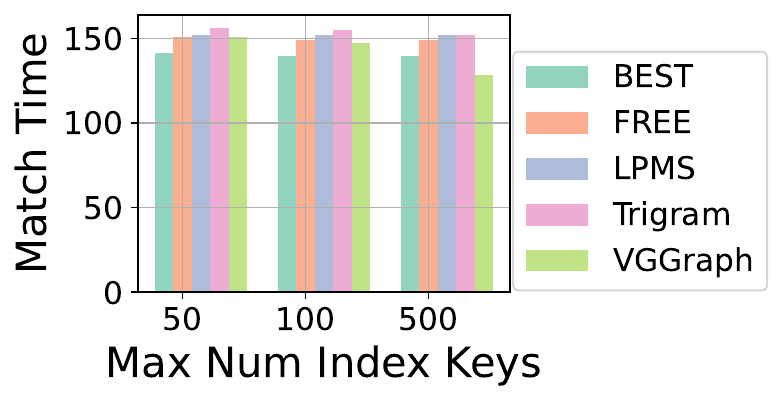}  
    \vspace{-0.5em}
  \caption{Query Time $T_Q$ (s)}
\label{fig:protein_queryTime}
\vspace{-1em}
\end{subfigure}
\vspace{-1.5em}
\caption{\wlprotein.\eat{ We use the bar plot to show the variable value of the highest-precision index for each method under different key number constraints $K$.}}
\label{fig:protein_perf}
\vspace{-1.5em}
\end{figure}

\best\ is the most performant for this workload, when $K$ is small, generating index with highest precision while using only 364 keys. The small alphabet size and small literal size per query make the number of possible n-grams considered in each calculation iteration small, and thus drastically reduce the index construction runtime overhead, which is the biggest advantage of \best\ in other workloads. {Similarly, \vgg\ which adapts a similar cover strategy, achieves the best precision as $K$ gets larger.}

{When indexing with 50 keys, \free($\max_n=2$,$c=0.15$) generates an index with a higher precision of 0.00651, while taking the least computational overhead and smallest index size.}
\lpms-D, taking slightly longer to select the n-grams and generate the index, achieves a precision of 0.00708 within 14 seconds.
\best($c=0.7$) achieves the highest precision, 0.00826, among the three methods. Its query time is also more than 10 seconds lower than the other two methods. {Overall index precisions are low for \wlprotein\ workload, as the average length of literals is very short in the regexes.}

When $K=100$, \free, \lpms, and \trigram\ does not lead to much improvement.
\eat{the index by \free($\max_n=2$,$c=0.2$) {has almost identical precision with the 50-key index constructed by the same method}. The workload running time for \lpms-D is slightly faster than \free.}
Index by \best($c=0.7$) has a higher precision of 0.0089, resulting in a more than 10 seconds workload performance improvement over other methods. \vgg\ ($max_n=6$, $c=0.7$) also increases slightly to 0.0732.
The precision increments for all methods does not corresponds to significant performance improvement as shown in the bars in~\Cref{fig:protein_queryTime}.\eat{ \sd{check the link}.}
As the key size constraint relaxes to $K=500$, selected n-grams of \vgg\ ($max_n=2$, $c=0.7$) achieves the highest precision of 0.0987, and the significant precision improvement from the previous $K$ led to a query time reduction to 128.5 seconds, $1.13\times$ query time reduction from the baseline. Since the literal lengths are small in this workload, the effectiveness of each n-gram is much more important for effective filtering. Therefore, n-gram sets selected by dataset-based methods, \free\ and \trigram, are not effective in filtering. 

\insight{For workloads where query literals are short and/or the alphabet size is small, \best\ can select an n-gram set with high filtering precision while incurring reasonable computational and storage overhead.  This advantage arises from the smaller n-gram candidate set. {\vgg\ outperform other methods in terms of precision as number of n-grams scales up.} \free\ does not benefit from this advantage since it only looks at n-grams in the dataset. \lpms\ benefits less from the characteristics of the workload due to the fixed overhead for constructing the integer program.
\eat{For workloads where query literals are short, workload-specific indexing or querying techniques that also consider the pattern components may further improve query performance, though that is beyond the scope of our work.}}

\eat{\best\ and \vgg\ approximate an optimal set cover when selecting n-grams, trades off n-gram selection time for higher precision n-grams tailored to the workload. The small alphabet size of \wlprotein\ reduces the number of candidate n-grams in calculation and therefore allow them to calculate a good n-gram set within reasonable time-frame. \lpms\ follows similar strategy, but its index construction time is still too long to finish within the set time limit, and we only get data points with index key number smaller than 100.
}

\begin{figure}[!tp]
\centering
\begin{subfigure}{0.43\columnwidth}
  \centering
  \includegraphics[width=0.95\linewidth]{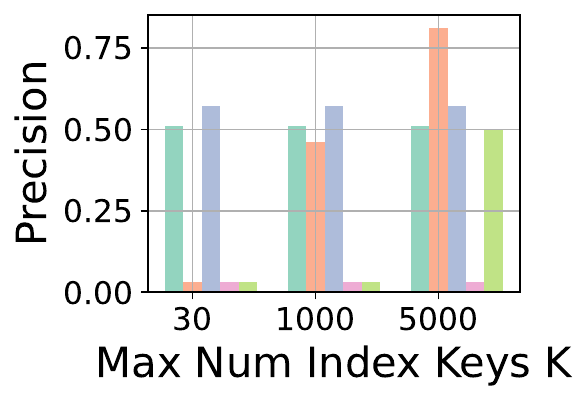}  
   \vspace{-0.5em}
  \caption{Precision}
  \label{fig:traffic_perc}
  \vspace{-1em}
\end{subfigure}
\begin{subfigure}{0.56\columnwidth}
  \centering
  \includegraphics[width=.95\linewidth]{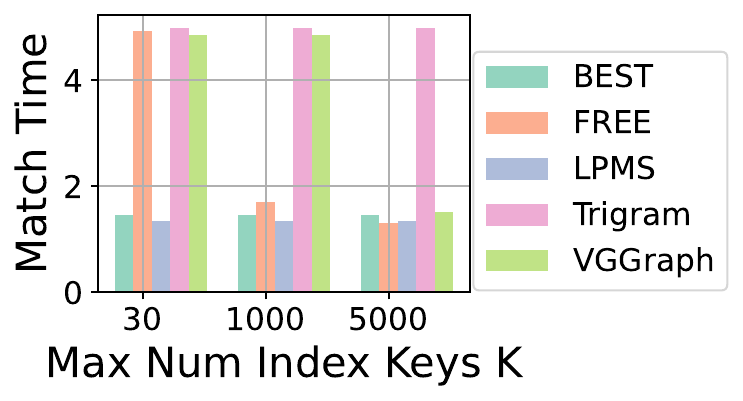}  
  \vspace{-0.5em}
  \caption{Query Time $T_Q$ (s)}
  \label{fig:traffic_queryTime}
  \vspace{-1em}
\end{subfigure}
\vspace{-1.5em}
\caption{\wltraffic.\eat{ We use the bar plot to show the variable value of the highest-precision index for each method under different key number constraints $K$.}}
\label{fig:traffic_perf}
\vspace{-1.5em}
\end{figure}

\subsubsection{Workload \wltraffic} This workload contains the smallest number of queries. Compared to the \wlweb\ workload, which also has a small query set, {the datasets and queries are generated by a limited number of templates}. The data records in the dataset consist of the location description (e.g., "At I-270, Between OH-48/Exit 29 and Dayton Intl Airport Rd/Exit 32") and a brief description of the accident. Each string in the \wltraffic\ dataset is much shorter. 

\Cref{fig:traffic_perf} shows the results. The methods that uses both query set and dataset as n-gram sources generate better query performance than the baseline query time of 4.8 seconds. 
\lpms-D performs the best for the \wltraffic\ workload, achieving a precision of 0.572 using only 12 keys. 
When $K$ increases to 1000, \free\ generates an index with 837 keys and a precision of 0.46, which is much lower than \best\ and \lpms\ with fewer than 30 keys. This is because \free\ selects n-grams based solely on their selectivity in the dataset, omitting those n-grams that are frequent in both the regex query and the dataset.
{When $K$ increases to 5000, \vgg\ ($max_n=3$, $c=0.7$) generates a index with 4247 keys and achieves an query time of 1.51 seconds that is comparable to other faster methods.
}
\eat{In fact, the configuration of \free\ that generates the highest precision with fewer than 1000 keys uses a selectivity threshold $c=0.7$, the largest across all of our experiments.}

\insight{For workloads where the query literals are very short and common in the dataset, \lpms\ can quickly locate a small set of n-grams with high precision. {\vgg,} \best, and \free\ remove n-grams that are frequent in the dataset, as they both adhere to Assumption~\ref{assumption:free:1}\eat{, an assumption} that does not hold for the dataset. \eat{Consequently, \best\ may eliminate beneficial n-grams due to the selectivity threshold, and \free\ may only index unnecessarily longer n-grams that benefit the queries.}
}

\begin{table}[t]
\centering
\captionof{table}{Index cost and query performance on \wlsql.\eat{ For each n-gram selection method under a specific key number constraint $K$ $s.t.$ $|I| \leq K$, we select the configuration with the highest precision. }}
 
\scalebox{0.9}{
\begin{tabularx}{\columnwidth}{ C|  c | R  R  R  R |  C}
\toprule
 $K$ & Method & $T_{Q}$ s & $T_{I}$ s &$S_{Q}$\scriptsize{GB} & $S_{I}$\scriptsize{MB} & $Prec$  \\ \midrule
\multirow{3}{*}{\bf{50}}& \free  & 3226.7 & 937 & \cellcolor{green!50}{76.7} & 6643.1 & 0.001480 \\ \cline{2-7}
& \lpms  & \cellcolor{green!50}{1421.1} & 4569 & 147.5 & \cellcolor{green!50}{4474.2} & \cellcolor{yellow!50}{0.009855} \\ \cline{2-7}
& \trigram  & 1552.3 & \cellcolor{green!50}{70} & 49.9 & 5529.5 & 0.003016 \\ \hline
\multirow{3}{*}{\bf{100}}& \free  & 3226.7 & 937 & \cellcolor{green!50}{74.0} & 6643.1 & 0.001480 \\ \cline{2-7}
& \lpms  & \cellcolor{green!50}{1421.1} & 4569 & 147.5 & \cellcolor{green!50}{4474.2} & \cellcolor{yellow!50}{0.009855} \\ \cline{2-7}
& \trigram  & 1552.3 & \cellcolor{green!50}{70} & 49.9 & 5529.5 & 0.003016 \\ 
\bottomrule
\end{tabularx}}
\vspace{-1em}
\label{tab:db_x_summary}
\end{table}

\subsubsection{Workload \wlsql} The \wlsql\ workload, like \wltraffic, consists of structured strings—specifically, system log reports—but with a much larger dataset and query set. \best\ was unable to complete within the time limit. Results are summarized in~\Cref{tab:db_x_summary}.

\lpms\ outperforms \free\ due to its query-aware design. With just 50 n-grams, \lpms-D achieves a precision of $9.855 \times 10^{-3}$ which is 10$\times$ higher than \free($\max_n=2$, $c=0.7$) under the same key limit. \free’s query-agnostic approach often selects short, selective n-grams from variable fields (e.g., VM IDs), which offer little benefit for filtering. Consequently, \lpms-D delivers a 2.43× faster query time. Although \lpms\ takes $3.4\times$ longer and requires $1.9\times$ more construction space and time overhead, the index built by \lpms\ uses $1.48\times$ less space. Since both methods use the same inverted index and number of keys, the difference stems from posting list lengths—indicating that \lpms\ selects more dataset-selective n-grams. This, combined with higher precision and faster queries, makes \lpms\ more effective overall. The results remain consistent until the upper limit of the number of n-grams grows to 5000. At this point, \free($\max_n=2$,$c=0.03$) selects an n-gram set of size 4643, increasing the precision to $2.68 \times 10^{-3}$. However, it still lags behind \lpms\ (even when using as low as $50$ n-grams) in both precision and query time.\eat{—despite similar index size and only slightly higher overhead.}

\insight{When query literals are common, the n-grams that have high \textit{benefit} may also have high \textit{selectivity}, which contradicts Assumption \ref{assumption:free:1}. For workloads where query literals are common in the dataset, even when the query literals are long, \lpms\ selects the n-gram set that improves query matching time the most. This is due to \lpms's design choice of not discarding any n-grams based on selectivity threshold.}

\subsubsection{Workload \wlenron\ } 
The \wlenron\ workload contains long data strings in the dataset like \wlweb, yet it has relatively small dataset size and larger query size. The lengths of literal components among queries are generally short. Although each data string is long, the small dataset size and the short literal components make the computation of majority of the methods within our set time limit. We run \best, \free, \lpms, \vgg, and \trigram\ on this dataset and summarize the results in~\Cref{tab:enron_summary}. Without indexing, \wlenron\ runs in 126 seconds as a baseline.

\best\ ($c=0.7$) achieves highest precision for $K=10$. With a precision of 0.148, around $1.4\times$ higher than other methods, \best\ uses $1.2\times$ less time for the workload querying. As $K$ increases to 100, precision of \best\ increases by $2.2\times$ to 0.321, and the query time decreases by $1.5\times$. \lpms-D, having similar strategy as \best, also achieves precision incremental as $K$ increases from 10 to 100 with significantly reduced precision. \wlenron\ has one of the highest regex match rate, where without filtering, the true positive rate per regex is already around 0.105, suggesting that \vgg\ and \trigram\ does not perform effective filtering.

When number of keys selected for the index continue to increase, the query performance and precision fails to increase proportionally. This is due to the fact that literal components in the queries are fairly common in the data strings, getting filtered out due to the selectivity threshold earlier on.

\insight{
{\best\ demonstrates exceptional performance in \wlenron, efficiently leveraging its optimization-based approach with smaller query sets and longer literals. \vgg\ shows substantial performance enhancement at moderate index sizes. \lpms\ offers balanced precision and performance, while \free\ and \trigram\ indexing lag behind due to their less targeted selection methods.}}

\begin{table}[t]
\centering
\captionof{table}{Index cost and query performance on \wlenron.\eat{ For each n-gram selection method under a specific key number constraint $K$ $s.t.$ $|I| \leq K$, we select the configuration with the highest precision. }}
\vspace*{-1em}
\scalebox{0.9}{
\begin{tabularx}{\columnwidth}{ C | C | R  R  R  R |  R }
\toprule
 $K$ & Method & $T_{Q}$ s & $T_{I}$ s &$S_{Q}$\scriptsize{GB} & $S_{I}$\scriptsize{MB} & $Prec$  \\ \midrule
\multirow{4}{*}{\bf{10}}& \best  & \cellcolor{green!50}{103} & 1781 & 3 & 5.7 & \cellcolor{yellow!50}{0.148} \\ 
& \free  & 126 & 30 & 3 & \cellcolor{green!50}{0.0} & 0.105 \\ 
& \lpms  & 124 & 79 & 3 & 16.3 & 0.107 \\ 
& \trigram  & 127 & \cellcolor{green!50}{4} & \cellcolor{green!50}{3} & 0.9 & 0.105 \\ \midrule
\multirow{5}{*}{\bf{100}}& \best  & \cellcolor{green!50}{69} & 2286 & 3 & 54.2 & \cellcolor{yellow!50}{0.321} \\ 
& \free  & 123 & 35 & 3 & 5.1 & 0.109 \\ 
& \lpms  & 101 & 2104 & 13 & 256.2 & 0.149 \\ 
& \trigram  & 127 & \cellcolor{green!50}{4} & \cellcolor{green!50}{3} & 0.9 & 0.105 \\ 
& \vgg  & 125 & 45 & 3 & \cellcolor{green!50}{0.4} & 0.105 \\ 
\bottomrule
\end{tabularx}}
\label{tab:enron_summary}
\vspace{-1em}
\end{table}

\subsubsection{Workload \wlsysy}
The \wlsysy\ workload has the largest dataset and a very small query set. \trigram\ method would not generate useful index if tight constraint is posed the number of n-grams selected, and will generate index too large to fit into memory if no constraint imposed. \vgg\ did not finish for all configuration on this workload. This workload runs in 1663 seconds without indexing. We run \free, \best, and \lpms\ on \wlsysy\ workload and summarize the result in~\Cref{tab:sysy_summary}.

Due to the small query set size, and large and diverse data strings in the dataset, methods considering both query and dataset, \best\ and \lpms\ achieves better precision and performance improvement for all key constraints $K$. Overall, \lpms-D achieves best precision of 0.328 and query performance of 221 seconds, $7.52\times$ faster than the baseline. \best\ ($c=0.1$) also achieves $5.3\times$ to $6.5\times$ performance improvement with 10 to 25 keys.

\insight{
{For \wlsysy, a workload with simple patterns and very large dataset size, \lpms\ stands out by effectively balancing precision and computational overhead. \best\ becomes computationally prohibitive due to its complexity, and \trigram\ indexing is impractical at this scale.}
}

\begin{table}[t]
\centering
\captionof{table}{Index cost and query performance on \wlsysy.\eat{ For each n-gram selection method under a specific key number constraint $K$ $s.t.$ $|I| \leq K$, we select the configuration with the highest precision. }}
\vspace*{-1em}
\scalebox{0.9}{
\begin{tabularx}{\columnwidth}{ C | C | R  R  R  R |  R }
\toprule
 $K$ & Method & $T_{Q}$ s & $T_{I}$ s &$S_{Q}$\scriptsize{GB} & $S_{I}$\scriptsize{MB} & $Prec$  \\ \midrule
\multirow{3}{*}{\bf{10}}& \best  & \cellcolor{green!50}{311} & 24810 & 217 & 351.5 & \cellcolor{yellow!50}{0.143} \\ 
& \free  & 1201 & \cellcolor{green!50}{1324} & \cellcolor{green!50}{43} & \cellcolor{green!50}{0.3} & 0.023 \\ 
& \lpms  & 373 & 2184 & 104 & 2799.0 & 0.113 \\ \midrule
\multirow{3}{*}{\bf{25}}& \best  & 254 & 28912 & 219 & 650.6 & 0.216 \\ 
& \free  & 1201 & \cellcolor{green!50}{1324} & \cellcolor{green!50}{43} & \cellcolor{green!50}{0.3} & 0.023 \\ 
& \lpms  & \cellcolor{green!50}{221} & 2174 & 105 & 3360.4 & \cellcolor{yellow!50}{0.328} \\ \midrule
\multirow{3}{*}{\bf{150}}& \best  & 254 & 28912 & 219 & \cellcolor{green!50}{650.6} & 0.216 \\ 
& \free  & 332 & \cellcolor{green!50}{947} & \cellcolor{green!50}{88} & 30360.0 & 0.189 \\ 
& \lpms  & \cellcolor{green!50}{221} & 2174 & 105 & 3360.4 & \cellcolor{yellow!50}{0.328} \\
\bottomrule
\end{tabularx}}
\label{tab:sysy_summary}
\vspace{-1em}
\end{table}

\subsection{Index Construction Overhead} \label{subsec:expr:index_overhead}

In this section, we aim to answer question~\ref{q:overhead}
with a detailed analysis of the index construction overhead for \free, \best, \lpms, \vgg, and \trigram, based on their time and space usage observed during index construction. 

\free\ demonstrates a significantly lower index construction time across various workloads due to its straightforward selectivity-based strategy and iterative prefix-free minimal n-gram selection approach. Specifically, for the \wlweb\ workload, as shown in~\Cref{tab:web_summary}, \free\ constructed an index with $1.7\times 10^5$ keys in merely 1184 seconds compared to \best's significantly longer 8587 seconds. \free's runtime space overhead remains relatively constant and low across workloads, generally proportional to the dataset size due to the prefix-free minimal set construction method.

\best\ has a high index construction overhead because it performs an exhaustive search to approximate optimal n-gram set through a utility-based set-cover strategy. Specifically, in the \wldblp\ workload, \best\ required 8762 seconds to construct an index of 363 keys, dramatically more than the sub-second construction time achieved by \free\ for a comparable number of keys. The high overhead results from the auxiliary data structures mentioned in~\Cref{method:best:approximate}, required to maintain for time-efficient computation.
In workloads with larger query sets or longer literals, such as \wlweb\ in~\Cref{tab:web_summary}, \best's computation overhead increases and potentially exceeding practical time limits.

\lpms\ has moderate to high index construction times, often higher than \free\ but lower than \best. For instance, in the \wlsql\ workload results in~\Cref{tab:db_x_summary}, \lpms\ required 4569 seconds for constructing an index with 50 keys, significantly exceeding \free’s 1342 seconds for an similar number of keys but less than \best. \lpms’s overhead results from its formulation and solving of the linear program.

\vgg\ has an index construction overhead generally lower than \best\ but higher than \free. The overhead is attributed to its iterative evaluation of cost-efficiency ratios. For the \wlsql\ workload in~\Cref{tab:db_x_summary}, \vgg\ constructed indices in 45 seconds, considerably lower than \best\ and \lpms\ that runs more than 2000 seconds but slightly higher than \free\ (30s). Its overhead scales as the dataset and query complexity increase. \trigram\ has minimal index construction overhead when limited to a small number of keys, but this overhead grows rapidly as more trigrams are indexed—especially without key constraints. Its cost is primarily driven by dataset size rather than query complexity or optimization.

\begin{figure}[!tp]
\centering
\begin{subfigure}{0.48\columnwidth}
  \centering
  \includegraphics[width=0.95\linewidth]{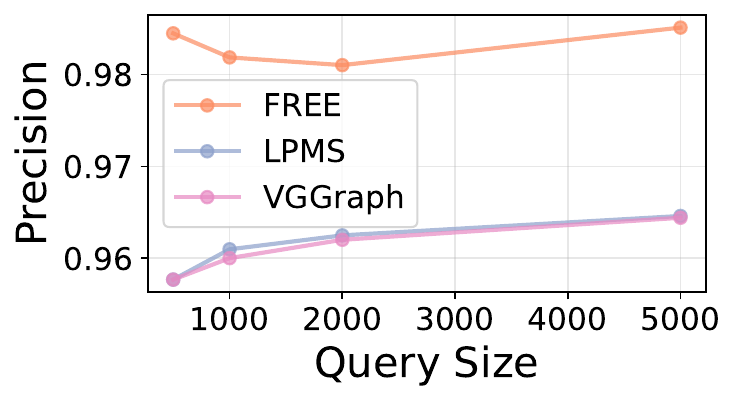}  
  \vspace{-0.5em}
  \caption{Precision}
  \label{fig:syn_prec_trend}
\end{subfigure}
\hfill
\begin{subfigure}{0.45\columnwidth}
  \centering
  \includegraphics[width=.95\linewidth]{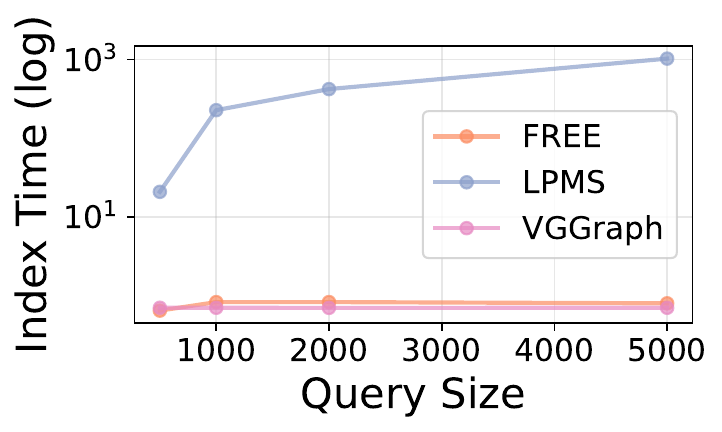}  
  \vspace{-0.5em}
  \caption{Index Time $T_I$ (in seconds)}
  \label{fig:syn_indexTime_trend}
\end{subfigure}
  \vspace{-0.5em}
\caption{\wlsynnew\ workload with $|D|=50,000$ and average query selectivity 0.02.\eat{ We use the bar plot to show the variable value of the highest-precision index for each method under different key number constraints $K$.}}
\label{fig:synthetic_trend}
\vspace{-1.5em}
\end{figure}

\subsection{Case Study: Workload Characteristics} \label{subsec:expr:workload_parameters}
In this section, we perform an ablation study with \wlsynnew\ workload to evaluate how different regex indexing methods behave under varying query set sizes and query selectivity. 
For each configuration, we constrain the maximum number of n-grams ($K$) to 5000 and measured both precision and query runtime ($T_Q$). \best\ failed to finish within the time limit.  However, in the experimental result, varying regex selectivity does not result in a common trend on the indices built with different methods in workload performance or index construction. We will focus on the analysis of varying query size with fixed dataset size and regex query selectivity. \eat{We plot our results in~\Cref{fig:synthetic_trend}.}

\Cref{fig:syn_prec_trend} illustrates that as the query set size increases from 1000 to 5000, \free\ consistently maintained high precision. This can be attributed to the small alphabet size of the workload. \lpms\ and \vgg\ also showed stable but lower precisions, with \lpms\ ranging around 0.95 and \vgg\ at approximately 0.96. All three methods achieves similarly high precision.
The precision of all methods slightly improve as query set size increased, as a larger query set provides more opportunities for n-gram filtering.
Regarding index construction time ($T_I$), in~\Cref{fig:syn_indexTime_trend}, \free\ and \vgg\ exhibited significantly lower overhead, maintaining fast index build times which is under 1 second across all query set sizes. \lpms\ required considerably more time as the query set size increased. \lpms\ is sensitive to query set growth, requires over 90 seconds at larger query sets. This matches our observation from~\Cref{subsec:expr:index_overhead}, where \lpms\ and \best\ incurred much larger index construction overhead.


\vspace*{1em}
\section{Learnings and Future Work}

\begin{figure}[!tp]
\centering
  \centering
  \includegraphics[width=0.85\linewidth]{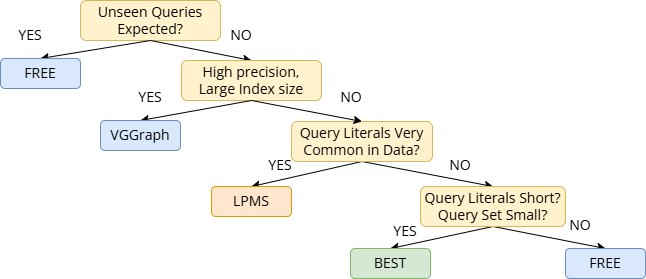} 
  \caption{N-gram selection and regex index method decision tree by workload characteristics.}
  \label{fig:dectree}
  \vspace*{-1em}
\end{figure}

The evaluation shows that the optimal choice of n-gram selection strategies among \free, \best, and \lpms\ is strongly influenced by workload characteristics (such as the query set size, query literal sizes, alphabet size, data patterns, etc.). Since no single strategy suits all workloads, this analysis emphasizes the importance of selecting methods based on workload characteristics and provides a general guideline. We summarize the insights in the decision flow chart in~\Cref{fig:dectree} as a guideline for practitioners. This guideline recommends 1) choosing \free\ for large and diverse workloads or when unseen queries are expected, 2) choosing \best\ when precision is crucial for repeated queries and the number of candidate n-grams is small, 3) choosing \lpms\ for formatted datasets or when query literals are common in the dataset, {and 4) choosing \vgg\ when precision requirement is high with low index space constraint.}

\eat{
\free\ is most suitable for large workloads, especially when query results are evenly distributed across the dataset \sd{why evenly distributed? what's the intuition?} \Ling{Ling: If that queries only focus on a very small subsets of the dataset, and other parts are totally different with no common chars, then the search space is effectively small, comparable to a smaller workload, where BEST might finish with reasonable time, and free would index a lot useless ngrams.}. It relies on the \textit{selectivity} of each individual n-gram for selection, ensuring efficient derivation of results. Therefore, it works best when the workload is too large for \best\ and \lpms. However, in workloads involving structured data (such as formatted logs), where frequent literals are essential to queries, its assumption that high-frequency n-grams are less \textit{useful} fails to hold.

\best\ prioritizes index precision and is optimal for small query sets, small alphabet sizes, or when the dataset is frequently searched. Its exhaustive search for a near-optimal n-gram set for indexing incurs a large computational overhead, rendering \best\ impractical for large workloads that would be queried very frequently. When the workload is run many times, the index construction overhead may be amortized. When the query set size or the alphabet size is small, \best's search space is reduced, and its exhaustive search for precision justifies the computational costs.

\lpms\ excels when query literals are highly common in the dataset, regardless of the dataset size, especially if we incorporate early stopping. The traditional selectivity assumption~\ref{assumption:free:1} held by \best\ and \free\ fails in workloads with frequently occurring query literals. \lpms\ approximates the results of \best\ with a more efficient algorithm, allowing it to skip the step in \best\ and \free\ that prunes n-grams based on individual selectivities. Therefore, \lpms\ can adapt to situations where highly beneficial n-grams are common in the dataset.
}

Based on this study, we also propose the following potential directions for future work in this area.

\begin{enumerate}[wide] 
    \item \textbf{Unified solution.} Our results demonstrate that while each method performs well on certain workloads, there is no solution that performs well across the board. Designing an algorithm that can combine the best aspects of each indexing method or prove the non-existence of such an algorithm would be beneficial. 
    \item \textbf{Better indexing formats.} Research in relation query processing has shown the benefits of using bit-based indexing formats (e.g. vector-based formats, BitWeaving~\cite{Li_2013}, etc.). It would be interesting to explore the benefits of those formats for regex indexing.
    \item \textbf{Indexing on-the-fly.} \looseness-1 All methods for regex indexing require an expensive preprocessing step. To the best of our knowledge, there are no existing indexing method that can perform indexing on-the-fly as the queries in the workload are executed and as the dataset and query distribution change dynamically. Adaptive query execution in this setting is another interesting direction for future work. 
\end{enumerate}

\vspace*{1em}
\section{Conclusion}

In this paper, we present an investigation of common regex indexing methods. Our comprehensive evaluation spans a diverse array of scenarios and datasets, establishing a benchmark for assessing the performance of these techniques. Through meticulous analysis, we have identified the inherent strengths and limitations of various methods. Based on our findings, we create a recommendation for practitioners on what method to choose based on the input characteristics. We also highlight the important open problems for regex indexing to spur further research in this area.


\balance
\bibliographystyle{ACM-Reference-Format}
\bibliography{references}

\appendix
\onecolumn

\begin{figure}[!tp]
\centering
\begin{subfigure}{0.5\linewidth}
  \centering
  \includegraphics[width=0.7\linewidth]{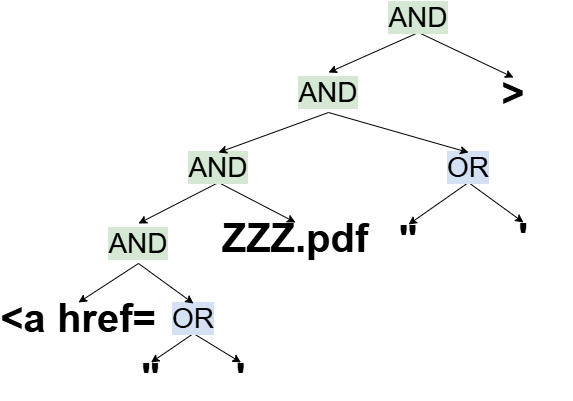}  
  \caption{Literal based query plan before checking index keys.}
  \label{fig:plan_tree_before}
\end{subfigure}
\hfill
\begin{subfigure}{0.46\linewidth}
  \centering
  \includegraphics[width=0.6\linewidth]{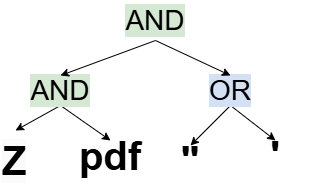}  
  \caption{Final query plan after index scan.}
  \label{fig:plan_tree_after}
\end{subfigure}
\caption{Example index search plan tree for regex \regexfmt{<a href=("|').*ZZZ\.pdf("|')>} built by \free, where the n-grams \regexfmt{"}, \regexfmt{'}, \regexfmt{Z}, and \regexfmt{pdf} are indexed.}
\label{fig:plan_tree}
\vspace*{-1em}
\end{figure}%

\section{\free\ Regex Query Plan Compiler}\label{app:free_plan}
In order to also handle regex queries literals in alternative strings, \free\ include a simple regex query plan compiler that compiles the overall index lookup to a plan tree with only logical \texttt{AND} and \texttt{OR} operators.
Specifically, for a given regex, the compiler identifies literals on the primary level and literals, and then build a tree with the literals. \free\ then checks if each of the literals or parts of the literals are indexed in the index, removing subtrees with no literals indexed, and substituting literal nodes with index keys, which are n-grams in the index.

We provide an example index search query plan in~\Cref{fig:plan_tree} for the example regex, \regexfmt{<a href=("|').*ZZZ\.pdf("|')>}, assuming that only \regexfmt{"}, \regexfmt{'}, \regexfmt{Z}, and \regexfmt{pdf} are indexed. The literal tree shown in~\Cref{fig:plan_tree_before} is built {solely} by extracting all literals from the regex. Referring to the index, we notice that \regexfmt{<a href=} and \regexfmt{>} are not indexed, and two n-grams in the literal \regexfmt{ZZZ.pdf}, \regexfmt{Z} and \regexfmt{pdf} are indexed. We simplify the plan to get~\Cref{fig:plan_tree_after}.

The evaluation of the index search plan tree then is a depth-first evaluation, where \texttt{AND} is a set intersection operation of two posting lists and \texttt{OR} is a set union operation of two posting lists.

\section{Robustness Test} \label{app:robustness}
For this test, we use the synthetic workload with different sets of regexes for index construction and query matching to test the robustness of each method for unseen queries. 

To test the robustness of the n-gram selection methods in handling unseen queries, we build off the synthetic dataset in the original \lpms~\cite{FAST} paper. The dataset consists of 5000 strings each constructed with alphabet `A-P'. The string size follows a geometric distribution with $p=1/32$. The index building query set is generated by randomly selecting 10\% from the dataset, and a random slice is used to create a regex \regexfmt{lit1 regex lit2}. \regexfmt{lit1} has 1-5 characters, \regexfmt{lit2} has 0-5 characters, and \regexfmt{regex} matches any $m$ characters where $1\leq m \leq 50$. During the query phase, we generate regexes of a similar format from 2\% of the data records. 

We summarize the result in~\Cref{tab:rob_summary}. Overall, \free\ demonstrates robustness in this experiment, achieving highest precision with the lowest computational overhead. This can be attributed to the fact that \free\ selects n-grams independent of the query set, and thus adapt smoothly to unseen regexes in query time.

For $K=20$, \free($\max_n=2$, $c=0.7$) builds an index with 16 keys that achieves the highest precision of 0.267. It also has the lowest index construction overhead. \best($c=0.2$) achieves slightly lower precision of 0.208, but the query time is $4.4\times$ more than that of \free. \lpms-D performs the worst with the lowest precision and highest query time. Note that \best\ and \lpms\ also have significantly higher computational overhead compared to \free.

As $K$ increases to 100, \best($c=0.5$) achieves the highest precision. It's index construction time also increases by $\approx$3 times to 31.7s. This is more than $1000\times$ longer than \free. Note that the query times for \best\ are not the lowest among the three methods, although it has the highest precision. This is due to the non-uniformness in the data record lengths, which breaks Assumption~\ref{assumption:best:2}.

\eat{As $K$ increases to 50 and 100, \best($c=0.7$) and \best($c=0.5$) achieves the highest precision within its group respectively. It's index construction time also increases from 3.4s when number of keys is 20 to 10.6s and 31.7s for $K=50$ and $K=100$ respectively. This is more than $1000\times$ longer than \free. Note that the query times for \best\ are not the lowest among the three although it has the highest precision. This is due to the non-uniformness in the data record lengths, which breaks Assumption~\ref{assumption:best:2}. }

As we increase to $K=300$, \free($max_n=2$, $c=0.12$) achieves the highest precision of 0.6453. \best($c=0.2$) achieves similar precision and similar overall query time compared to \free, but with a much larger index construction time which is $2833\times$ slower than \free. In all cases, \lpms\ performs unfavorable to the other two methods, since the n-gram selection strategy of \lpms\ is solely based on the benefit calculation using the query set and the dataset, without any heuristic like the pruning step of \best.

\insight{
For scenarios when a unseen queries on a fixed dataset are expected, the strategy that takes advantage of the query set (e.g. \best\ and \lpms) may fall short. \free, selecting n-grams based on the dataset, is robust in this setting. \eat{When the query set is a representative sample of the dataset, or when we have an exhaustive n-gram set with a loose index size constraint, indices built by \best\ can also achieve comparably high precision. However, it comes with a cost of a several magnitude slower index construction overhead. }
}

\begin{table}[t]
\centering
\captionof{table}{Index cost and query performance on \wlsyn. }
 
\scalebox{0.9}{
\begin{tabularx}{\columnwidth}{ C|  C | R  R  R  R |  R }
\toprule
 $K$ & Method & $T_{Q}$ s & $T_{I}$ s &$S_{Q}$\scriptsize{GB} & $S_{I}$\scriptsize{MB} & $Prec$  \\ \midrule
\multirow{3}{*}{\bf{20}}& \best  & 0.829 & 3.398 & 0.042 & 0.091 & 0.2084 \\ 
& \free  & \cellcolor{green!50}{0.190} & \cellcolor{green!50}{0.027} & \cellcolor{green!50}{0.016} & 0.446 & \cellcolor{yellow!50}{0.2672} \\ 
& \lpms  & 1.211 & 3.720 & 0.053 & \cellcolor{green!50}{0.010} & 0.1479 \\ \midrule
\multirow{3}{*}{\bf{100}}& \best  & 0.444 & 31.659 & 0.043 & 0.452 & \cellcolor{yellow!50}{0.4793} \\ 
& \free  & \cellcolor{green!50}{0.190} & \cellcolor{green!50}{0.027} & \cellcolor{green!50}{0.016} & 0.446 & 0.2672 \\ 
& \lpms  & 1.063 & 3.791 & 0.052 & \cellcolor{green!50}{0.016} & 0.1757 \\ \midrule
\multirow{3}{*}{\bf{300}}& \best  & \cellcolor{green!50}{0.061} & 90.661 & 0.043 & 1.125 & 0.6451 \\ 
& \free  & 0.064 & \cellcolor{green!50}{0.032} & \cellcolor{green!50}{0.015} & 1.122 & \cellcolor{yellow!50}{0.6453} \\ 
& \lpms  & 1.063 & 3.791 & 0.052 & \cellcolor{green!50}{0.016} & 0.1757 \\ 
\bottomrule
\end{tabularx}}
\label{tab:rob_summary}
\end{table}

\end{document}